\documentclass{PoS}
\usepackage{amssymb,amsmath,mathptmx,bm,bbm}
\usepackage{epsf}
\usepackage{epsfig,rotate}
\usepackage{afterpage}
\usepackage{longtable}
\usepackage{cite}
\usepackage{color}

\let\normalcolor\relax

\newcommand{\vus}{|V_{us}|}
\newcommand{\vcb}{|V_{cb}|}

\newcommand{\vub}{|V_{ub}|}
\newcommand{\vts}{|V_{ts}|}

\def\epe{\varepsilon'/\varepsilon}
\newcommand{\tev}{\, {\rm TeV}}
\newcommand{\gev}{\, {\rm GeV}}
\newcommand{\mev}{\, {\rm MeV}}

\newcommand{\mw}{M_{\rm W}}

\newcommand{\be}{\begin{equation}}
\newcommand{\ee}{\end{equation}}
\newcommand{\bea}{\begin{eqnarray}}
\newcommand{\eea}{\end{eqnarray}}

\newcommand{\ord}{{\cal O}}

\def\kpn{K^+\rightarrow\pi^+\nu\bar\nu}

\def\klpn{K_{L}\rightarrow\pi^0\nu\bar\nu}

\title{Flavour Theory: 2009
}
\ShortTitle{Flavour Theory: 2009}

\author{\speaker{Andrzej J. Buras}\\
Technical University Munich, Physics Department, D-85748 Garching, Germany,\\
 TUM Institute for Advanced Study, D-80333 M\"unchen, Germany \\
   E-mail: \email{aburas@ph.tum.de}}

\abstract{
After an overture and a non-technical exposition
of the relevant theoretical framework including a brief discussion 
of some of the most popular extensions of the Standard Model,
we will compile a list of 20 goals in flavour physics that could be 
reached already in the next decade.
In addition to $K$, $D$ and $B_{s,d}$ decays and lepton flavour
violation  also flavour conserving observables like 
electric dipole moments of the neutron and leptons and $(g-2)_\mu$ are
included in this list. Flavour violation in high energy processes 
is also one of these goals.
Subsequently we will discuss in more detail some of the most urgent issues for
the coming years in the context of 
several extensions of the Standard Model like models with Minimal Flavour
Violation, the general MSSM, the Littlest Higgs Model with T parity, 
Randall-Sundrum models and supersymmetric flavour models.
This presentation is not meant
to be a comprehensive review of flavour physics but rather a personal 
view on this fascinating field and an attempt to collect  those routes
that with the help of upcoming experiments should allow us to reach
a much deeper understanding of  physics, in particular flavour physics,
 at very short distance scales. 
}

\FullConference{European Physical Society Europhysics Conference on 
   High Energy Physics,
EPS-HEP 2009,\\
		 July 16 - 22 2009\\
		 Krakow, Poland}

\begin{document}

\section{Overture}
\label{sec:intro}
The main goal of elementary particle physics is to search for physics laws
at very short distance scales. From the Heisenberg uncertainty principle 
\cite{Heisenberg:1927nn}  we
know that to test scales of order $10^{-18}{\rm m}$ we need the energy of 
approximately $200\gev$. With approximately $E=4\tev$,
effectively available at the LHC, we will be able to probe distances as short
as $5\cdot 10^{-20}{\rm m}$. Unfortunately, it is unlikely that we can do
better before 2046 through high energy collider experiments. On the other
hand flavour-violating and CP-violating processes are very strongly suppressed
and are governed by quantum fluctuations that allow us to test energy scales 
as high as $200\tev$ corresponding to short distances in the ballpark of
$10^{-21}{\rm m}$. Even shorter distance scales
can be tested, albeit indirectly, in this manner. Consequently frontiers
in testing ultrashort distance scales belong to flavour physics or more
concretely to very rare processes like particle-antiparticle mixing, rare decays of mesons, CP violation
and lepton flavour violation. Also electric dipole moments and $(g-2)_\mu$ 
belong to these
frontiers even if they are flavour conserving.
While such tests are not limited by the available energy, they are limited
by the available precision. The latter has to be very high as the 
Standard Model (SM) has been until now very successful and finding 
departures from its predictions has become a real challenge. 

Flavour physics developed over the last two decades into a very broad field.
In addition to $K$, $D$ and $B_d$ decays and $K^0-\bar K^0$ and $B_d-\bar B_d$
mixings that were with us for quite some time, $B_s^0-\bar B_s^0$ mixing, 
$B_s$ decays and $D^0-\bar D^0$
mixing belong these days to the standard repertoire of any flavour 
workshop. Similarly lepton flavour violation (LFV) 
gained in importance after the
discovery of neutrino oscillations and related non-vanishing neutrino 
masses even if within the SM the LFV is basically unmeasurable. 
Simultaneously new ideas for the explanation of the quark and lepton
mass spectra and the related weak mixings, summarized by the
CKM \cite{Cabibbo:1963yz,Kobayashi:1973fv} and PMNS 
\cite{Pontecorvo:1957qd,Maki:1962mu}
matrices, developed significantly in this decade. Moreover the analyses
of electric dipole moments (EDM's), of the $(g-2)_\mu$ anomaly and of 
flavour changing neutral current (FCNC) processes in top quark decays 
intensified during the last years in
view of the related experimental progress that is expected to take place
in the next decade.

The correlations between all these observables and the interplay of
flavour physics with direct searches for new physics (NP)
 and electroweak precision
studies will tell us hopefully one day which is the proper extension of the
SM.

In preparing this talk I have been guided by the impressive success of the CKM
picture of flavour changing interactions 
\cite{Cabibbo:1963yz,Kobayashi:1973fv}, evident in the excellent talks of
Adrian Bevan \cite{Bevan:2009qe} and Giovanni Punzi, and also by several tensions between the
flavour data and the SM that possibly are the first signs of NP. Fortunately,
there is still a lot of room for NP contributions, in particular in rare 
decays of mesons and charged leptons, in CP-violating transitions and in
electric dipole moments of leptons, of the neutron and of other particles. 
There
is also a multitude of models that attempt to explain the existing tensions
and to predict what experimentalists should find in the coming decade. 
Yet, in my opinion, those models should be favoured at present that try to
address the important open questions of contemporary particle physics
like the issue of the stabilization of the Higgs mass under loop corrections 
and the question of the origin of the observed hierarchies in fermion 
masses and mixings. Such extensions will play the dominant role in this 
report.

There is also the important question whether the footprints of NP that 
is responsible for the hierarchies in question will be seen directly at
the LHC and indirectly through flavour and CP-violating processes in the
coming decade. Hoping that this is indeed the case we will assume in what
follows that the NP scales in various extensions of the SM discussed 
below are not larger that $2-3\tev$, so that the new particles predicted 
by these extensions are  in the reach of the LHC.

After a brief recollection of the theoretical framework and the 
description of the most popular NP scenarios in Section 2, we will list
in Section 3, the twenty most important goals in this field for the
coming decade. There is no space to discuss all these goals in detail here.
Therefore in  Section 4 we will only discuss  the 
ones which in my opinion are the most important at present. A number of
enthusiastic statements will end this report.

I should strongly emphasize that I do not intend to present here a totally 
comprehensive review 
of flavour physics. Comprehensive reviews, written by a hundred of 
flavour experts are already present on the market
\cite{Buchalla:2008jp,Raidal:2008jk,Antonelli:2009ws} and moreover,
extensive studies of the physics at future flavour machines and other visions
can be found in \cite{Bona:2007qt,Browder:2008em}.
  I would rather like to paint a picture of flavour physics in 
general terms and collect various strategies for the exploration of this fascinating 
field that hopefully will turn out to be useful in the coming years. In this context
 I will recall present puzzles in flavour 
physics that could turn out to be the first hints of NP and on various
occasions I will present the predictions of the NP scenarios mentioned in the
Abstract.
Last but certainly not least let me cite two excellent text books on CP 
violation and flavour physics \cite{Branco:1999fs,Bigi:2000yz}, where many fundamentals of this field are
clearly explained and other extensions of the SM and other observables are
discussed in detail.

\section{Theoretical Framework}
\subsection{Preliminaries}

The starting point of any serious analysis of weak decays in 
the framework of a 
given extension of the SM is the basic Lagrangian 
\be\label{basicL}
{\cal L} ={\cal L}_{\rm SM}(g_i,m_i,V^{i}_{\rm CKM})+
{\cal L}_{\rm NP}(g_i^{\rm NP},m_i^{\rm NP},
V^{i}_{\rm  NP}),
\ee
where $(g_i,m_i,V^{i}_{\rm CKM})$  denote the parameters of the SM and
$(g_i^{\rm NP},m_i^{\rm NP},V^{i}_{\rm  NP})\equiv\varrho_{\rm NP}$
 the additional 
parameters in a given NP scenario.

Our main goal then is to identify in weak decays the effects  decribed by  
${\cal L}_{\rm NP}$
 in the presence of 
the background from ${\cal L}_{\rm SM}$. 
In the first step one derives the Feynman rules following 
from (\ref{basicL}), which allows to calculate Feynman diagrams. But then we
 have to face two challenges:

\begin{itemize}
\item
our theory is formulated in terms of quarks, but experiments involve their 
bound states: $K_L$, $K^\pm$, $B_d^0$, $B_s^0$, $B^\pm$, $B_c$, $D$, $D_s$, 
etc. 
\item
NP takes place at very short distance scales $10^{-19}-10^{-18}~{\rm m}$, 
while $K_L$, $K^\pm$, $B_d^0$, $B_s^0$, $B^\pm$ and other mesons live at
$10^{-16}-10^{-15}~{\rm m}$.
\end{itemize}

The solution to these challenges is well known. 
One has to construct an effective 
theory relevant for experiments at low energy scales. 
Operator Product Expansion (OPE)
and Renormalization Group (RG) methods are involved here. 
They allow to separate 
the perturbative short distance (SD) effects, where NP is present, 
from long distance 
(LD) effects for which non-perturbative methods are necessary. Moreover RG 
methods allow an efficient summation of large logarithms 
$\log (\mu_{\rm SD}/\mu_{\rm LD})$. 
A detailed exposition of 
these techniques can be found in \cite{Buchalla:1995vs,Buras:1998raa}
 and fortunately we do not have to 
repeat them here. At the 
end of the day the formal expressions involving matrix elements of local operators 
and their Wilson coefficients can be cast into the following {\it Master Formula for 
Weak Decays} \cite{Buras:2001pn}.

\subsection{Master Formula for Weak Decays}

 The master formula in question reads:
\be\label{master}
{\rm A(Decay)}=\sum_i B_i \eta^i_{\rm QCD}V^i_{\rm CKM} 
F_i(m_t,{\rm \varrho_{NP}}),
\ee
where $B_i$ are non-perturbative parameters representing hadronic matrix
elements of the contributing operators, $\eta_i^{\rm QCD}$ stand symbolically 
for the renormalization group factors, $V^i_{\rm CKM}$ denote the relevant
combinations of the elements of the CKM matrix and finally 
$F_i(m_t,\varrho_{\rm NP})$
denote the loop functions resulting in most models from box and penguin
diagrams 
but in some
models  also representing tree level diagrams if such diagrams contribute.
The internal charm contributions have been suppressed in this formula but
they have to be included in particular in $K$ decays and $K^0-\bar K^0$ 
mixing.
$\varrho_{\rm NP}$ denotes symbolically all parameters beyond $m_t$, in 
particular the set $(g_i^{\rm NP},m_i^{\rm NP},V^{i}_{\rm  NP})$ in 
(\ref{basicL}).
It turns out to be useful to factor out $V^i_{\rm CKM}$ in all contributions
in order to see transparently the deviations from Minimal Flavour Violation 
(MFV).

In the SM only a particular set of parameters $B_i$ is relevant as there are
no right-handed charged current interactions, the
functions $F_i$ are {\it real} and the flavour and CP-violating effects
 enter only through 
the CKM factors $V^i_{\rm CKM}$.
 This  implies that the
functions $F_i$ are universal with respect to flavour so that they are the
same in the $K$, $B_d$ and $B_s$ systems. Consequently a number of observables 
in these different systems are strongly correlated with each other 
within the SM.

 The simplest class of extensions of the SM are  models with 
Constrained Minimal Flavour Violation (CMFV)
\cite{Buras:2000dm, Buras:2003jf,Blanke:2006ig,Buras:2009us}.
They are formulated as follows:
\begin{itemize}
\item
All flavour changing transitions are governed by the CKM matrix with the 
CKM phase being the only source of CP violation,
\item
The only relevant operators in the effective Hamiltonian below the weak scale
are those that are also relevant in the SM.
\end{itemize}
This implies that relative to the SM  only the values of $F_i$ are 
modified but their universal character remains intact. In particular they
are real. Moreover,
 in cases where 
$F_i$ can be eliminated by taking certain combinations of observables,
universal correlations between these observables
 for this class of models result. We will encounter some of these correlations
in Section 4.

In more general MFV models 
\cite{D'Ambrosio:2002ex,Chivukula:1987py,Hall:1990ac}
new parameters $B_i$ and 
$\eta_i^{\rm QCD}$, related to new operators, enter the game but the functions
$F_i$ still remain real quantities as in the CMFV
framework and do not involve any flavour violating parameters. Consequently
the CP and flavour violating effects in these models 
are again governed by the CKM matrix.
However, the presence of new operators makes this approach less constraining
than the CMFV framework. We will discuss some other aspects of this approach 
below.

In the simplest non-MFV models, the basic operator structure of CMFV models
remains but the functions $F_i$ in addition to real SM contributions can
contain new flavour parameters and new complex phases. Consequently the
CKM matrix ceases to be the only source of flavour and CP violation.

Finally, in the most general non-MFV models, new operators 
(new $B_i$ parameters)
contribute and the functions $F_i$ in addition to real SM contributions can
contain new flavour parameters and new complex phases.

 Obviously this classification of different classes of models corresponds 
to a $2\times 2$ matrix but before presenting this matrix 
let us briefly discuss  the essential ingredients in our master formula. 

 Clearly without a good knowledge of  non-perturbative factors $B_i$ 
no precision studies of 
flavour physics will be possible unless the non-perturbative uncertainties 
can be reduced or even removed by taking suitable ratios of observables. 
In certain rare 
cases it is also possible to measure the relevant hadronic 
matrix elements entering 
rare decays by using leading tree level decays. Examples of such fortunate 
situations are certain mixing induced CP asymmetries and 
the branching ratios for 
$K\to\pi\nu\bar\nu$ decays. Yet, in many cases one has to face the direct
evaluation of
$B_i$. While 
lattice calculations, QCD-sum rules, Light-cone sum rules and large-$N$ methods made
significant progress in the last 20 
years, the situation is clearly not satisfactory and one should hope that new 
advances in the calculation of $B_i$ parameters will be made in the LHC era 
in order to 
adequately use improved data. Recently an impressive progress in 
calculating the 
parameter $\hat B_K$, relevant for CP violation in $K^0-\bar K^0$ mixing,
 has been made and we will discuss its implications in Section 4.

An important progress has also been made in organizing the dominant
contributions in non-leptonic two-body $B$ meson decays and decays like 
$B\to V\gamma$ with the help of the QCD factorization approach, SCET and
the Perturbative QCD approach.

Concerning the factors $\eta^i_{\rm QCD}$ 
 an
impressive progress has been made during the last 20 years. The 1990's can be 
considered as the era of NLO QCD calculations. Basically the NLO corrections to all 
relevant decays and transitions have been calculated already 
in the last decade \cite{Buchalla:1995vs},  
with a few exceptions, like the width differences $\Delta\Gamma_{s,d}$ in
the $B^0_{s,d}-\bar B^0_{s,d}$ systems that were completed only in 2003
\cite{Beneke:1998sy,Beneke:2002rj,Ciuchini:2003ww}. This 
decade can be considered as the era of NNLO calculations. 
In particular one should 
mention here the NNLO calculations of QCD corrections to $B\to X_sl^+l^-$
\cite{Asatryan:2002iy,Asatrian:2002va,Gambino:2003zm,Ghinculov:2003qd,Ghinculov:2002pe,Bobeth:2003at,Beneke:2004dp},
 $\kpn$ \cite{Buras:2005gr,Buras:2006gb,Brod:2008ss},
and 
in particular to $B_s\to X_s\gamma$ \cite{Misiak:2006zs}
  with the latter one being by far 
the most difficult one. Also important steps towards a complete calculation 
of NNLO corrections to non-leptonic decays 
of mesons have been made in \cite{Gorbahn:2004my}.

The final ingredients of our master formula, in addition to $V^i_{\rm CKM}$
factors, are the loop functions $F_i$ resulting from penguin
and
 box diagrams with the exchanges of the
top quark, $W^\pm$, $Z^0$, heavy new gauge bosons, heavy new fermions and
scalars. They are known at one-loop level in 
several extensions of the SM, in particular in the two Higgs doublet model
(2HDM), the
littlest Higgs model without T parity (LH), the ACD model with one universal
extra dimension (UED), the MSSM with MFV and non-MFV violating interactions,
the flavour blind MSSM (FBMSSM), the littlest Higgs model with T-parity (LHT),
  $Z^\prime$-models, Randall-Sundrum (RS) models,
left-right
symmetric models, the model with the sequential fourth generation of 
quarks and 
leptons. Moreover, in the SM $\ord(\alpha_s)$ corrections
to all relevant one loop functions are known. It should also be stressed
again that
in the loop functions in our master formula one can conveniently absorb tree
level FCNC contributions present in particular in RS models.

After this symphony of names like FBMSSM, LH, LHT, RS let us explain them
briefly by summarizing the most popular extentions of the SM.

\subsection{Minimal Flavour Violation}
We have already formulated what we mean by CMFV and MFV.
 Let us first add here that  the models with CMFV generally contain only
one Higgs doublet and  the top Yukawa coupling dominates.
On the other hand models with MFV in which 
the operator structure differs from
the SM one contain  two Higgs doublets and bottom and top Yukawa 
couplings can be of comparable size. A well known example is the MSSM with MFV
and large $\tan\beta$. The MFV framework can be elegantly formulated 
with the help of global symmetries present in the limit of vanishing 
Yukawa couplings \cite{Chivukula:1987py,Hall:1990ac} and its implications 
can be studied efficiently with the help of spurion 
technology \cite{D'Ambrosio:2002ex,Feldmann:2006jk}. However, I will
not enter this presentation here as it can be found in basically any 
paper that discusses MFV. Recent discussions of various aspects of MFV 
can be found in 
\cite{Colangelo:2008qp,Paradisi:2008qh,Mercolli:2009ns,Feldmann:2009dc,Kagan:2009bn,Paradisi:2009ey}.

Here let us only stress that the MFV symmetry principle in itself does 
not forbid the presence of
{\it flavour blind} CP violating sources~\cite{Baek:1998yn,Baek:1999qy,Bartl:2001wc,Ellis:2007kb,Colangelo:2008qp,Altmannshofer:2008hc,Mercolli:2009ns,Feldmann:2009dc,Kagan:2009bn}.
Therefore, in particular,  a MFV MSSM suffers from the same SUSY CP problem 
as the ordinary MSSM.
Either an extra assumption or a mechanism accounting for a natural suppression
of these CP-violating
phases is desirable.
The authors of~\cite{D'Ambrosio:2002ex} proposed the extreme situation where the SM Yukawa couplings
are the only source of CPV. In contrast, recently in~\cite{Paradisi:2009ey}, 
such a strong assumption has been
relaxed and the following generalized MFV ansatz has been proposed: the SUSY breaking mechanism is
{\it flavour blind} and CP conserving and the breaking of CP only arises through the MFV compatible
terms breaking the {\it flavour blindness}. That is, CP is preserved by the sector responsible for
SUSY breaking, while it is broken in the flavour sector.
While the generalized MFV ansatz still accounts for a natural solution of the SUSY CP problem, it also
leads to peculiar and testable predictions in low energy CP violating processes~\cite{Paradisi:2009ey}.

The MFV approach is simple and offers an elegant explanation of the fact
that the CKM framework works so well even if NP is required
to be present at scales $\ord(1\tev)$. But one has to admit that 
it is a rather pessimistic approach to NP. The deviations
from the SM expectations in CP conserving processes amount  in the 
case of CMFV to at most
$50\%$ at the level of the branching ratios 
\cite{Bobeth:2005ck,Haisch:2007ia,Hurth:2008jc}. More generally in the MFV 
framework only in cases where scalar 
 operators
are becoming important and helicity suppression in decays like 
$B_s\to \mu^+\mu^-$ is lifted, enhancements of the relevant branching 
ratios by more than a factor of two and even one order of
magnitude relative to the SM are  possible. However, independently of
whether it is CMFV or MFV, the CP violation in this class of models 
is SM-like and in order to be able to
distinguish among various models in this class high precision will be required
which calls for experiments like Super-Belle, Super-B facility in Frascati
 and $K\to\pi\nu\bar\nu$ experiments like NA62 and KOTO.

One should also emphasize that MFV in the quark sector does not
offer the explanation of the size of the observed baryon-antibaryon 
asymmetry in the universe (BAU) and it does not address the hierarchy 
problem related to the quadratic divergences in the Higgs mass.
Similarly the hierarchies in the quark masses and  quark mixing angles remain
unexplained in this framework. For this reason there is still potential
interest in non-MFV new physics scenarios to which we will now turn our
attention.

\subsection{Most Popular Non-MFV Extensions of the SM}
The search for NP at the $1\tev$  scale is centered already for 
three decades 
around the hierarchy problem, be it the issue of quadratic divergences in 
the Higgs 
mass, the disparity of the electroweak, GUT and Planck scales or 
the doublet-triplet 
splitting in the context of SU(5) GUTs. The three 
 most popular 
directions which aim to solve at least some of these problems are as follows:

{\bf a) Supersymmetry (SUSY)}

In this approach the cancellation of quadratic divergences in $m_H$ is
achieved with the help of new particles of different spin-statistics than 
the SM particles:
supersymmetric particles.
 For this approach to 
work, these new particles should have masses below $1\tev$, 
otherwise the fine 
tuning of parameters cannot be avoided. One of the important predictions of 
the 
simplest realization of this scenario, the MSSM with R-parity, 
is the light Higgs with 
$m_H\le 130\gev$ and one of its virtues is its perturbativity up to the GUT
scales. 

The ugly feature of the General MSSM (GMSSM) 
is a large number of parameters residing 
dominantly in the soft sector
that has to be introduced in the process of supersymmetry breaking. 
Constrained 
versions of the MSSM can reduce the number of parameters significantly. 
The same is true in the case of the MSSM with MFV. An excellent review on 
supersymmetry can
be found in \cite{Martin:1997ns}.

The very many new flavour parameters in the
soft sector makes the GMSSM  not very predictive and 
moreover this framework is plagued by flavour and CP problems:
FCNC processes and EDM's are generically well above
the experimental data and upper bounds, respectively. Moreover the GMSSM
framework addressing primarily the gauge hierarchy problem and the
quadratic divergences in the Higgs mass does not provide automatically the
hierarchical pattern of quark and lepton masses and of the hierarchical
pattern of their FCNC and CP-violating interactions.

Much more interesting from this point of view are supersymmetric flavour
models (SF) with flavour symmetries that allow a simultaneous understanding of 
the flavour structures in the Yukawa couplings and in SUSY soft-breaking 
terms, adequately suppressing FCNC and CP-violating phenomena and solving
SUSY flavour and CP problems. A recent detailed study of various SF models 
has been performed in \cite{Altmannshofer:2009ne}.
We have analysed there 
the following representative scenarios in which NP contributions are characterized
by:
\begin{itemize}
\item [i)] The dominance of right-handed (RH) currents 
(abelian model by Agashe and Carone\cite{Agashe:2003rj}),
\item [ii)] Comparable left- and right-handed currents with CKM-like mixing
  angles represented by the special version (RVV2) 
of the non abelian $SU(3)$ 
model by
Ross, Velasco and Vives \cite{Ross:2004qn} as discussed recently in \cite{Calibbi:2009ja} and 
the model by Antusch, King and Malinsky (AKM) \cite{Antusch:2007re},
\item [iii)] The dominance of left-handed (LH) currents in non-abelian 
models~\cite{Hall:1995es} ($\delta$LL) .
\end{itemize}
Through a model-independent analysis we have found that  
these three scenarios
predicting  quite 
different patterns of flavour violation should give a good representation of
most SF models discussed in the literature. Short summaries of our results 
can be found in \cite{Altmannshofer:2009ap,Buras:2009hr}. 

In Section 4 we will mainly confine our presentation
of predictions of supersymmetry  
to these SF models. However, we will also briefly encounter the MSSM with 
MFV in which new 
{\it flavour blind} but {CP-violating} phases are present. This FBMSSM 
framework has
been discussed in 
\cite{Baek:1998yn,Baek:1999qy,Bartl:2001wc,Ellis:2007kb} and last year
in \cite{Altmannshofer:2008hc}, where a number of correlations between
various flavour conserving and flavour violating observables, both
CP-violating, has been pointed out.

Next, let us recall that the new particles in supersymmetric models,
that is  squarks, sleptons, gluinos, charginos,
 neutralinos, charged Higgs particles $H^{\pm}$ and additional 
 neutral scalars can contribute to FCNC processes
through virtual exchanges in box and  penguin diagrams.
Moreover, new sources of flavour
and CP violation come from the misalignement of quark and squark mass matrices
 and similar new flavour and CP-violating effects are present in the lepton
sector. Some of these effects can be strongly enhanced at large $\tan\beta$.
Finally, in the MSSM a useful parametrization of the new effects is
given by $\delta_{ij}^{AB}$ with $i,j=1,2,3$ and $A ,B=L,R$
 in the context of the so-called mass insertion approach 
\cite{Hall:1985dx,Gabbiani:1996hi}. However, it should be emphasized that 
 in certain models, 
like supersymmetric flavour models, this approximation is not always 
accurate and exact diagonalization of squark mass matrices is mandatory 
in order to obtain meaningful results \cite{Dedes:2008iw,Altmannshofer:2009ne}.

{\bf b)	Little Higgs Models}

In this approach the cancellation of divergences in $m_H$ is achieved with 
the help of new particles of the same spin-statistics. 
Basically the SM Higgs is kept light 
because it is a pseudo-Goldstone boson of a new spontaneously broken global 
symmetry. 
Thus the Higgs is protected by a global symmetry from acquiring a large mass, 
 although in order to achieve this the weak gauge group has to be extended and 
the Higgs mass generation properly arranged 
({\it collective symmetry breaking}). The 
dynamical origin of the global symmetry in question and the physics behind its 
breakdown is not specified. But in analogy to QCD one could imagine 
a new strong 
force at scales $\ord(10\tev)$ among  new very heavy fermions that bind 
together to produce the SM Higgs. In this scenario the SM Higgs is 
analogous to 
the pion. At 
scales well below $10\tev$ the Higgs is considered as an 
elementary particle but at
$10\tev$  its 
composite structure should be seen.  At these high scales one will have to 
cope with non-perturbative strong  dynamics, and an unknown ultraviolet 
completion with some impact on 
low energy predictions of 
Little Higgs models has to be specified. 
The advantage of these models, relative to supersymmetry, is a 
much smaller number of free parameters. Excellent reviews can be found in
\cite{Schmaltz:2005ky,Perelstein:2005ka}.

In Little Higgs models in contrast to the MSSM, new heavy gauge bosons $W_H^\pm$, $Z_H$ and
 $A_H$  in the case of the so-called littlest Higgs model 
without \cite{Arkani-Hamed:2002qy} and with T-parity 
\cite{Cheng:2003ju,Cheng:2004yc} are expected. Restricting our 
discussion to the model with T-parity (LHT), the masses of $W_H^\pm$ and
 $Z_H$ are typically $\ord(700\gev)$. $A_H$ is significantly lighter with
 a mass of a few hundred GeV and being the lightest particle 
with odd T-parity can play the role of the dark matter candidate. 
Concerning the 
fermion sector, there is a new very heavy $T$-quark necessary to cancel 
the quadratic 
divergent contribution of the ordinary top quark to $m_H$ and a copy of all SM
quarks and leptons is required by T-parity. These mirror quarks and mirror 
leptons 
interact with SM particles through the exchange of
$W_H^\pm$, $Z_H$ and  $A_H$ gauge bosons which in turn implies 
 new flavour and CP-violating contributions to decay amplitudes that are 
governed by  new mixing matrices
in the quark and lepton sectors. These matrices
 can have very different structure than the CKM and PMNS matrices.
The mirror quark and leptons can have masses in the range 
of 500-1500\gev and could be discovered at the LHC. As we will see in Section 4 
their impact on
FCNC processes can be sometimes spectacular.
Reviews on flavour physics in 
the LHT model can be found in 
\cite{Blanke:2007ww,Duling:2007sf,Blanke:2009am}.

{\bf c)	Extra Space Dimensions}

When the number of space-time dimensions is increased,
 new solutions to the hierarchy 
problems are possible. Most 
ambitious proposals are models with a warped extra dimension first proposed
by Randall and Sandrum (RS)  \cite{Randall:1999ee} which provide a geometrical
explanation of the
hierarchy  between the Planck scale and the EW scale. Moreover, when the SM
fields, except for the Higgs field, are
allowed to propagate in the bulk 
\cite{Gherghetta:2000qt,Chang:1999nh,Grossman:1999ra}, 
these models naturally generate the
hierarchies in the fermion masses and mixing angles 
\cite{Grossman:1999ra,Gherghetta:2000qt} through different localisations 
of the fermions in the bulk. Yet, this way of explaining the hierarchies in masses 
and mixings necessarly   
implies FCNC transitions
at the tree level 
\cite{Burdman:2003nt,Huber:2003tu,Agashe:2004cp,Csaki:2008zd}.
The RS-GIM mechanism 
\cite{Huber:2003tu,Agashe:2004cp}, combined with an additional custodial
protection of  flavour violating $Z$ couplings 
\cite{Blanke:2008zb,Blanke:2008yr,Buras:2009ka}, 
 allows yet to achieve 
the agreement with
existing data without a considerable fine tuning of parameters.
 Reviews of \cite{Blanke:2008zb,Blanke:2008yr,Buras:2009ka}
 can be found in
 \cite{Duling:2009sf,Gori:2009tr,Buras:2009us,Blanke:2009mn,Duling:2009vc,Gori:2009em,Buras:2009hr}. 
New theoretical ideas addressing the issue of large FCNC transitions in the
RS framework and proposing new protection mechanisms occasionally leading
to MFV can be found in 
\cite{Csaki:2008eh,Cacciapaglia:2007fw,Cheung:2007bu,Santiago:2008vq,Csaki:2009bb,Csaki:2009wc}.

In extra dimensional models obvious signatures in high energy processes 
are the lightest Kaluza-Klein particles, the
excited sisters and brothers of
the SM particles that  can also have significant impact on low energy 
processes. When KK-parity is present, like in models 
with universal extra 
dimensions, then also a dark matter candidate is present. In models with
warped extra dimensions and protective custodial symmetries 
\cite{Agashe:2003zs,Csaki:2003zu,Agashe:2006at,Blanke:2008zb,Blanke:2008yr} 
imposed to avoid problems
with electroweak precision tests (EWPT) and the data on FCNC processes, 
the gauge group is
generally larger than the SM gauge group and similar to the LHT model
new heavy gauge bosons are present. However, even in models with custodial
symmetries these gauge bosons must be sufficiently heavy ($2-3\tev$) in order 
to be consistent with EWPT. We will denote such RS framework with custodial 
symmetries by RSc.

As far as the 
gauge boson sector of the RSc model is concerned, 
in addition to the SM gauge bosons
the lightest new gauge bosons are the KK--gluons, the KK-photon and the 
electroweak KK gauge bosons $W^\pm_H$, $W^{\prime\pm}$, $Z_H$ and $Z^\prime$,
all with masses $M_{KK}$ around $2-3\tev$. The fermion sector is 
enriched through heavy KK-fermions (some of them with exotic electric charges)
 that could in principle be discovered at 
the LHC. The fermion content
of this model is explicitly given in \cite{Albrecht:2009xr}, where also 
 a complete set of 
Feynman rules has been worked out. Detailed analyses of electroweak precision
tests and of the parameter $\varepsilon_K$ in a RS model without custodial 
protection can be found in \cite{Casagrande:2008hr,Bauer:2008xb}. 

{\bf d)	Other Models}
 
 There are several other models studied frequently in the literature. 
These are in particular $Z^\prime$ models and models with vector-like 
heavy quarks \cite{delAguila:2000rc,Branco:2006wv,Picek:2008dd}. 
Both are present in the RS scenario and I will not discuss
them separately. Recently new interest arose in models with a sequential 
 4th generation 
which is clearly  a possibility. In particular George Hou 
\cite{Hou:2008di,Hou:2008yb,Hou:2008ji}
 and subsequently 
Lenz \cite{Bobrowski:2009ng}, 
Soni \cite{Soni:2008bc}
and their collaborators made extensive analyses of FCNC processes in
this framework. See also \cite{Eilam:2009hz}. This NP scenario is quite different from SUSY, the LHT and RS 
models as the 4th generation of quarks and leptons cannot decouple and if these new 
fermions 
exist, they will be found at the LHC. However this direction by itself 
does not address any hierachy problems and I will not further discuss 
it in this 
report. Electroweak precision tests in the presence of fourth generation and 
other constraints 
are discussed in 
\cite{Fok:2008yg,Chanowitz:2009mz,Novikov:2009kc,Burdman:2008qh}.

\subsection{The Flavour Matrix}
The discussion of Section 2.2 suggests to exhibit different extensions of 
the SM in  form
of a $2\times 2$ matrix  shown in Fig.1. Let us briefly describe the four
entries of this matrix.

The element (1,1) or the class A represents models with 
CMFV. The SM, the 
versions of 2HDM's with low $\tan\beta$, the LH model 
and the ACD model \cite{Appelquist:2000nn}
 with a universal fifth flat extra dimension 
belong to this class. 
\begin{figure}
\begin{center}
\vspace*{-0.5cm}
\includegraphics[width=3in,angle=270]{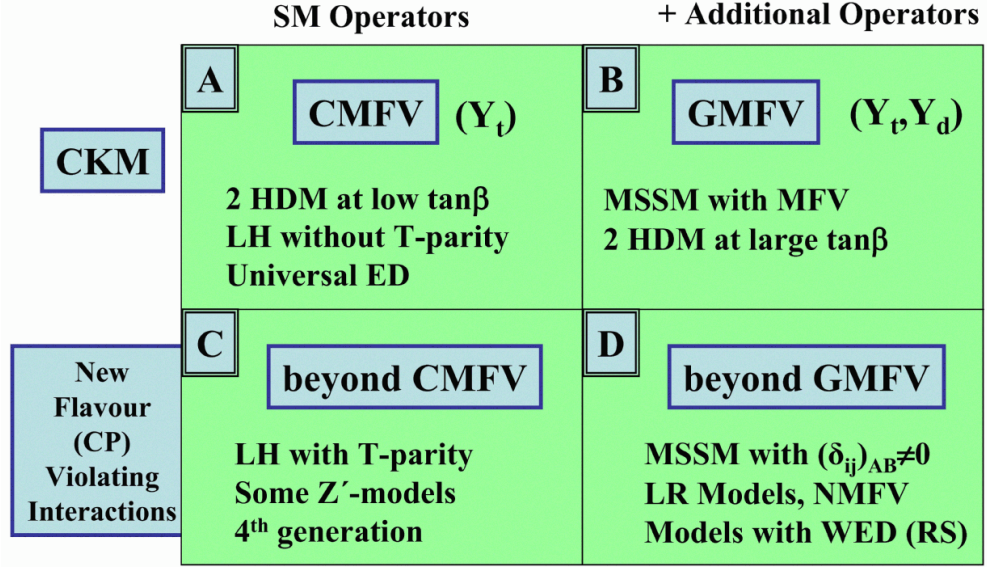}
\vspace*{-0.5cm}
\end{center}
\caption{\label{fig:matrix}The Flavour Matrix}
\end{figure}

The elements (1,1) and (1,2) or classes A and B taken together, the upper 
row of the flavour matrix, 
represent the class of models with MFV at large.  
Basically the new effect in the
(1,2) entry  relative to 
(1,1) alone is the appearance of new operators with different 
Dirac structures that 
are strongly suppressed in the CMFV framework but can be enhanced if $\tan\beta$ is large or 
equivalently if $Y_d$ cannot be neglected. 
2HDM with large $\tan\beta$ belongs to this class. 
In the past it was believed that the MSSM  corresponds to the 
entry (1,2) only with large $\tan\beta$ but the analysis in \cite{Altmannshofer:2007rj}
 has shown that even at low
$\tan\beta$
 $Y_d$ cannot be 
neglected when the parameter $\mu$ in the Higgs sector is large and gluino
contributions  become important. We will see below that the presence of new
operators,  in particular scalar 
operators, allows to lift the helicity suppression of certain rare decays 
like $B_s\to\mu^+\mu^-$, 
resulting in very different predictions than found in the CMFV models.

The FBMSSM scenario carrying new complex phases that are flavour conserving
represents a very special class of MFV models in which the functions $F_i$
become complex quantities in contrast to what we stated previously but 
as these new phases are flavour conserving a natural place for FBMSSM is
the upper row of the flavour matrix.

A very interesting class of models is the one represented by the entry 
(2,1) or the class C. Relative to CMFV it contains new flavour violating interactions, in
particular new complex phases, forecasting novel
 CP-violating effects that may significantly differ from those present in the CMFV class. 
As there are no new operators relative to the SM ones, 
no new $B_i$-factors and 
consequently no new non-perturbative uncertainties relative to CMFV models are 
present. Therefore predictions of models belonging to the (2,1) entry suffer generally 
from smaller non-perturbative uncertainties than models represented by 
the second 
column in the flavour matrix in Fig.~1.

When discussing the (2,1) models, it is important to distinguish between models in 
which new physics couples dominantly to the third generation of quarks, basically 
the top quark, and models where there is a new sector of fermions that can 
communicate with the SM fermions with the help of new gauge interactions. 
Phenomenological approaches with enhanced Z-penguins 
\cite{Buras:1998ed,Buras:1999da,Buras:2004ub}, 
some special $Z^\prime$-models \cite{Langacker:2000ju,Barger:2009qs,Langacker:2009im} and
 the fourth generation models
\cite{Hou:2008di,Soni:2008bc,Bobrowski:2009ng,Eilam:2009hz}
belong to the first subclass of (2,1), while the
LHT model represents the second subclass.

The entry (2,2) represents the most complicated class of models 
in which both new flavour violating effects and new operators are relevant. 
The MSSM with flavour violation coming from the squark sector and RS 
models
are likely to be  the most 
prominent members of this class of models.
The NMFV approach of \cite{Agashe:2005hk} and
 left-right symmetric models belong also 
to this class. The spurion 
technology for this class of models has been developed by Feldmann and
Mannel \cite{Feldmann:2006jk}. 

\subsection{The Little Hierarchy Problem}

As we have seen, the stabilization of the Higgs mass under radiative 
corrections 
requires NP at scales $\ord(1\tev)$. Yet EWPT 
performed first at LEP/SLC and subsequently 
extended at Tevatron imply that NP, unless 
properly screened, can only appear at scales of 5-10 TeV or higher. 
The situation is 
much worse in FCNC processes. There the masses of new particles carrying
 flavour and having $\ord(1)$ couplings cannot contribute at tree level 
unless their masses are 
larger than $1000\tev$ or even more. A detailed analysis of this issue can be
found in particular in \cite{Bona:2007vi}.

Thus in order to keep the solutions to the hierarchy problems
 discussed above alive, 
protective symmetries must be present in order to suppress NP effects to 
electroweak precision observables (EWPO) 
and to FCNC processes in spite of NP being present at scales
 $\ord(1 \tev)$ or 
lower. In this context the custodial SU(2) symmetry in the case of EWPT 
should be mentioned. In the framework of the LHT model this 
symmetry is guarantied by  T-parity. For the FCNC processes 
we need generally a GIM mechanism which forbids tree level
contributions. If this mechanism is violated and FCNC transitions occur
already at tree level other protections are necessary. In RS models the
so-called RS-GIM mechanism \cite{Huber:2003tu,Agashe:2004cp}
and the recently pointed
out custodial protection for flavour violating 
$Z$ couplings \cite{Blanke:2008zb,Blanke:2008yr,Buras:2009ka} play an 
important role.

In this context MFV
is very popular as models with MFV can naturally
satisfy the existing FCNC constraints.
While this framework  will play a role below, we will in Section 4
dominantly 
present the results coming from the non-MFV scenarios discussed in 
Section 2.4.

\section{20 Goals in Flavour Physics for the Next Decade}
We will now list 20 goals in flavour physics for the coming decade. 
The order in which these goals will be listed does not represent by any 
means a ranking in importance. In this section each goal will be summarized 
very briefly including some references where further details can be found. 
In Section 4 we will concentrate on the goals 1, 3, 4, 6 and 10 which most 
likely will play the central role in  quark flavour physics in the 
coming years. We will close Section 4 by correlating these goals with 
the goals 16, 17 and 18 that deal with lepton physics in the context of 
supersymmetric flavour models. Let us now list the 20 goals in question.

\newpage

{\bf Goal 1: The CKM Matrix from tree level decays}

This determination would give us the values of the elements of the CKM 
matrix without NP pollution. From the present perspective most important are 
the determinations of
$\vub$ and $\gamma$ because they are presently not as well known as
$\vcb$ and $\vus$. However, a precise determination of $\vcb$ is also
important as $\varepsilon_K$, $Br(\kpn)$ and $Br(\klpn)$ are roughly 
proportional to $\vcb^4$. While Super-B facilities accompanied by improved 
theory should be able to determine $\vub$ and $\vcb$ with precision of
$1-2\%$, the best determination of the angle $\gamma$ in the first half
of the next decade will come from LHCb. An error of a few degrees should
be achievable around 2015 and this measurement could also be improved 
at Super-B machines.

{\bf Goal 2: Improved Lattice Calculations of Hadronic Parameters}

The knowledge of  the meson decay constants $F_{B_s}$, $F_{B_d}$ and of various
$B_i$ parameters with high precision would allow in conjunction with Goal 1
to make precise calculations of $\Delta M_s$, $\Delta M_d$, $\varepsilon_K$, 
$Br(B_{s,d}\to\mu^+\mu^-)$ and of other observables in the SM. We  could 
then directly see whether the SM is capable
of describing these observables or not. The most recent unquenched calculations allow
for optimism and in fact a very significant progress in the calculation of 
$\hat B_K$ has been made recently. We will discuss its implications in 
Section 4.

For completeness we collect here some selected non-perturbative parameters 
relevant for FCNC processes.
The present lattice values, that are relevant for $B_{s,d}^0-\bar B^0_{s,d}$ 
mixings, taken from
 \cite{Lubicz:2008am} read
\be
F_{B_s} \sqrt{\hat B_{B_s}} = 270(30)\mev,\qquad
F_{B_d} \sqrt{\hat B_{B_d}} = 225(25)\mev,
\ee
while the HPQCD collaboration \cite{Gamiz:2009ku} finds similar values 
but smaller errors,
 \be
F_{B_s} \sqrt{\hat B_{B_s}} = 266(18)\mev,\qquad
F_{B_d} \sqrt{\hat B_{B_d}} = 216(15)\mev.
\ee

Other values that should be improved are the $\hat B_i$ parameters themselves
that will play some role in predicting the branching ratios for 
$B_{s,d}\to \mu^+\mu^-$ as we proceed. The present lattice results read
\cite{Lubicz:2008am}
\begin{equation}\label{BBB}
\frac{\hat B_{s}}{\hat B_{d}}=1.00\pm 0.03, \qquad
\hat B_{d}=1.22\pm0.12, \qquad \hat B_{s}=1.22\pm0.12~.
\end{equation}
Also the accuracy of the $B_i$ parameters related to new operators present in 
the classes $B$ and $D$ in the flavour matrix should be improved.

In this context one should mention the determination of quark masses 
and of the QCD coupling constant $\alpha_s(M_Z)$ that should still be improved
in order to reduce the parametric uncertainties in the predictions for
various branching ratios. 
Here important advances have been made recently. Let 
me just quote \cite{Chetyrkin:2009fv}
\be\label{mc}
m_b(m_b)= (4.163\pm0.016)\gev,
\qquad m_c(m_c)=1.279\pm0.013\gev, 
\ee
with the latter very relevant for the decay $\kpn$.
Similarly,
\be
m_s(2\gev)= (91 \pm7)\mev, 
\qquad m_t(m_t)=163.5\pm1.7\gev~,
\ee
with the value of $m_s(2\gev)$ given recently by Leutwyler 
\cite{Leutwyler:2009xx}. This agrees very well with  
 \cite{Jamin:2006tj}, where $94\pm6\mev$ has been quoted.

Finally, two impressive determinations of $\alpha_s(M_Z)$ should be 
mentioned here. One is from hadronic Z and $\tau$ decays \cite{Baikov:2008jh} 
 resulting
in $\alpha(M_Z)=0.1198\pm0.0015$ and the second from the $\tau$ hadronic 
width \cite{Beneke:2008ad} with the result $\alpha(M_Z)=0.1180\pm0.0008$.
 The latest  world average reads \cite{Bethke:2009jm}
\be
\alpha(M_Z)=0.1184\pm0.0007~.
\ee

{\bf Goal 3: Is $\mathbf{\varepsilon_K}$ consistent with
 $\mathbf{S_{\psi K_S}}$ within the SM?}

The recent improved value of $\hat B_K$ from unquenched lattice QCD 
acompanied by a more careful look at $\varepsilon_K$ suggest that
the size of CP violation measured in $B_d\to\psi K_S$ might be insufficient
to describe $\varepsilon_K$ within the SM. Clarification of this new tension 
is important as the $\sin 2\beta-\varepsilon_K$ correlation in the SM is presently the
only relation between CP violation in the $B_d$ and $K$ systems that can be 
tested experimentally. We will return to this issue in Section 4.

{\bf Goal 4: Is $\mathbf{S_{\psi\phi}}$ much larger than its tiny SM value?}

Within the SM CP violation in the $B_s$ system is predicted to be very small. 
The best known representation of this fact is the value of the mixing 
induced CP asymmetry: $(S_{\psi\phi})_{\rm SM}\approx 0.04$. The present data from
CDF and D0 indicate that CP violation in the $B_s$ system could be much
larger, $S_{\psi\phi}=0.81^{+0.12}_{-0.32}$ \cite{Barberio:2007cr}. 
This is a very interesting deviation from the SM. 
Its clarification is of utmost importance and I will return to this question
in Section 4. Fortunately, we should know the answer to this question within the
coming years as CDF, D0, LHCb, ATLAS and CMS will make big efforts to 
measure $S_{\psi\phi}$ precisely.

{\bf Goal 5: Non-Leptonic Two-Body B Decays and Related Puzzles}

The best information on CP violation in the $B$ system to date comes 
from two-body non-leptonic decays of $B_d$ and $B^\pm$ mesons. While
until now these decays dominated this field, LHCb 
will extend these studies in an important manner
 to $B_s$ and $B_c$ decays. This is clearly
a challenging field not only for experimentalist but in particular also for
theorists due to potential hadronic uncertainties. Yet, in the last 
ten years an impressive progress has been made in measuring many 
channels, in particular $B\to\pi\pi$ and $B\to\pi K$ decays, and in
developing a number of methods like QCD factorization 
\cite{Beneke:1999br,Beneke:2000ry}, the Perturbative QCD
approach \cite{Keum:2002vi}, SCET
\cite{Bauer:2001yt,Bauer:2002nz,Bauer:2002uv,Beneke:2002ni,Beneke:2002ph} and
more phenomenological approaches based on 
flavour
symmetries \cite{Buras:2004ub,Chiang:2004nm}. Excellent reviews of this subject have been given 
by   Buchalla \cite{Buchalla:2008tg}, Fleischer \cite{Fleischer:2008uj}
 and Silvestrini \cite{Silvestrini:2007yf}. They contain a lot
of useful material. I think this field will
continue to be important for the tests of the CKM
framework in view of very many channels whose branching ratios should be
measured in the next decade with a high precision. This is also a place 
where the structure of QCD effects in the interplay with weak interactions
can be studied very well and the combination of the lessons gained from
this field with those coming from theoretically cleaner decays 
discussed subsequently will undoubtly enrich our view on flavour 
physics. 

On the other hand in view of potential
hadronic uncertainties present in the branching ratios and direct CP 
asymmetries these observables 
in my opinion will  not provide definite answers about
NP if the latter contributes to them only at the level of $20\%$ or less. 
On the 
other hand mixing induced CP-asymmetries like $S_{\psi K_S}$, 
$S_{\psi\phi}$ and alike being theoretically much cleaner 
will continue to be very important for the tests of NP.
Let me then  just briefly discuss a number of departures from
the SM predictions which await resolution in the coming years.

First of all
the angle $\beta$ has been measured in several other decays, in particular in
penguin dominated decays like $B\to \phi K_S$ or $B\to \eta^\prime K_S$ with
the result that it is generally smaller than $(\sin 2\beta)_{\psi K_S}$, 
putting the SM and MFV in some difficulties.
Clarification of this disagreement is an important goal for the next decade.
While this tension became weaker with time, the theoretically clean
asymmetry $S_{\phi K_S}$ still remains to be significantly smaller than
the expected value of approximately $0.67$ \cite{Barberio:2007cr}:
\be\label{spK}
S_{\phi K_S}=0.44\pm 0.17.
\ee
This tension cannot be resolved at LHCb and its resolution will remain
as  one of the 
important goals for Super Belle at KEK and later the Super-B machine in 
Frascati, 
although an insight on a possible anomalous behaviour in this asymmetry 
should be gained 
at  LHCb through the study of CP violation in $B_s\to\phi\phi$ 
\cite{Fleischer:2007wg}.

We will see in Section 4
 that the desire to explain the value in (\ref{spK}) in the 
framework of some supersymetric  models will
have interesting implications for other CP-violating observables like
the direct CP asymmetry in $B\to X_s\gamma$ and electric dipole moments.

Next
the rather large difference
in the direct CP asymmetries $A_{CP}(B^{-}\to K^{-}\pi^{0})$ and
$A_{CP}(\overline{B}^{0}\to K^{-}\pi^{+})$ observed by the Belle and BaBar 
collaborations has not been  expected but it could be due to
our
insufficient understanding of hadronic effects rather than NP. 
Similar comments apply to certain puzzles 
in $B\to\pi K$ decays \cite{Buras:2004ub} which
represent additional tensions that decreased with time but did not
fully disappear \cite{Baek:2009hv}. For a different view see 
\cite{Ciuchini:2008eh}. 

Finally of particular interest is the mixing induced CP-asymmetry in
$B\to\pi^0K_S$ which appears to indicate still some tensions with the
 SM expectations
\cite{Buras:2004ub,Fleischer:2008wb,Gronau:2008gu} although this 
is inconclusive at present.
For the most recent analysis see \cite{Ciuchini:2008eh}.

{\bf Goal 6: $\mathbf{Br(B_{s,d}\to\mu^+\mu^-)}$}

In the SM and in several of its extentions  $Br(B_{s}\to\mu^+\mu^-)$ 
is found in the ballpark of $3-5\cdot 10^{-9}$, which is by an order of
magnitude lower than the present bounds from CDF and D0. A discovery of 
$Br(B_{s,d}\to\mu^+\mu^-)$ at $\ord (10^{-8})$ would be a clear signal of
NP, possibly related to Higgs penguins. LHCb can reach the SM level for 
this branching ratio in the first years of its operation. From my point of 
view, similar to $S_{\psi\phi}$, precise measurements of 
$Br(B_{s}\to\mu^+\mu^-)$ and later of a more suppressed branching ratio
$Br(B_{d}\to\mu^+\mu^-)$ are among the most important goals in flavour 
physics in the coming years. We will discuss both decays in Section 4.

{\bf Goal 7:  
$\mathbf{B\to X_{s,d}\gamma}$, $\mathbf{B\to K^*(\varrho)\gamma}$ and 
$\mathbf{A_{\rm CP}^{\rm dir}(b\to s\gamma)}$}

The radiative decays in question, in particular $B\to X_s\gamma$, played 
an important role in constraining NP in the last 15 years because both 
the experimental data and also the theory have been already in a good
shape for some time with the NNLO calculations of $Br(B\to X_s\gamma)$ 
being at the forefront of perturbative QCD calculations in weak decays.
Both theory and experiment reached roughly $10\%$ precision and the 
agreement of the SM with the data is good implying not much 
room left for NP contributions. Still further progress both in theory and 
experiment should be made to further constrain NP models.
This will only be possible when Super-B machines enter their operation. 
Of particular interest is the
direct CP asymmetry $A_{\rm CP}^{\rm dir}(b\to s\gamma)$ that is 
similar to $S_{\psi\phi}$ predicted to be tiny ($0.5\%$) in the 
SM but
could be much larger in some of its extensions as discussed in Section 4.

\newpage

{\bf Goal 8: $\mathbf{B\to X_sl^+l^-}$ and
                     $\mathbf{B\to K^*l^+l^-}$}

While  the branching ratios for $B\to X_sl^+l^-$ and  $B\to K^*l^+l^-$
put already significant constraints on NP, the angular observables, 
CP-conserving ones like the well known forward-backward asymmetry 
and CP-violating ones will definitely be very useful for distinguishing
various extensions of the SM. Recently, a number of detailed analyses 
of various CP averaged symmetries and CP asymmetries provided by the 
angular distributions in the exclusive decay $B\to K^*(\to K\pi)l^+l^-$
have been performed in 
\cite{Bobeth:2008ij,Egede:2008uy,Altmannshofer:2008dz}.
In particular the zeroes of some of these 
observables can be accurately predicted. Belle and BaBar provided already
interesting results for the best known forward-backward asymmetry but 
the data have to be improved in order to see whether some sign of NP 
is seen in this asymmetry. Future studies by the LHCb and Super-B machines
will be able to contribute here in a significant manner.

{\bf Goal 9: $\mathbf{B^+\to \tau^+\nu}$ and
                     $\mathbf{B^+\to D^0\tau^+\nu}$}

The SM expression for the branching ratio of the tree-level decay $B^+ \to \tau^+ \nu$ is given by
\begin{equation} \label{eq:Btaunu}
{Br}(B^+ \to \tau^+ \nu)_{\rm SM} = \frac{G_F^2 m_{B^+} m_\tau^2}{8\pi} \left(1-\frac{m_\tau^2}{m^2_{B^+}} \right)^2 F_{B^+}^2 |V_{ub}|^2 \tau_{B^+}~.
\end{equation}
In view of the parametric uncertainties induced in (\ref{eq:Btaunu}) by $F_{B^+}$ and $V_{ub}$,
in order to find the SM prediction for this branching ratio one can 
rewrite it as follows \cite{Altmannshofer:2009ne}:
\begin{eqnarray} \label{eq:Btaunu_DMd}
{Br}(B^+ \to \tau^+ \nu)_{\rm SM} &=& \frac{3 \pi}{4
    \, \eta_B \, S_0(x_t) \, \hat B_{B_d}} \frac{m_\tau^2}{M_W^2} \left(1 -
  \frac{m_\tau^2}{m_{B^+}^2} \right)^2 \left\vert \frac{V_{ub}}{V_{td}}
  \right\vert^2 \tau_{B^+}~\Delta M_d ~.
\end{eqnarray}
 Here $\Delta M_d$ is supposed to be taken from experiment and
\begin{equation}
\left\vert \frac{V_{ub}}{V_{td}}\right\vert^2 =
\left( \frac{1}{1-\lambda^2/2} \right)^2
~\frac{1+R_t^2-2 R_t\cos\beta}{R^2_t} ~,
\end{equation}
with $R_t$ and $\beta$ determined by means of (\ref{eq:Rt_sin2beta_SM}) in 
Section 4.
In writing (\ref{eq:Btaunu_DMd}),
we used $F_B \simeq F_{B^+}$ and $m_{B_d}\simeq m_{B^+}$. We then find
\cite{Altmannshofer:2009ne}
\begin{equation}\label{eq:BtaunuSM1}
{Br}(B^+ \to \tau^+ \nu)_{\rm SM}= (0.80 \pm 0.12)\times 10^{-4}.
\end{equation}
 This result
agrees well with a recent result presented by the UTfit collaboration
\cite{Bona:2009cj}. 

On the other hand, the present experimental world avarage based 
on results by BaBar \cite{Aubert:2007xj,:2008gx} and Belle
\cite{Ikado:2006un,:2008ch} reads \cite{Tisserand:2009ja}
\begin{equation} \label{eq:Btaunu_exp}
{Br}(B^+ \to \tau^+ \nu)_{\rm exp} = (1.73 \pm 0.35) \times 10^{-4}~,
\end{equation}
which is roughly by a factor of 2 higher than the SM value.
We can talk about a tension at the $2.5\sigma$
level.
 
While the final data from BaBar and Belle will lower the exparimental
error on  $Br(B^+\to\tau^+\nu)$, the full clarification of a possible
discrepancy between the SM and the data will have to wait for the
data from Super-B machines. Also improved values for $F_B$ from lattice 
and $\vub$ from tree level decays will be important if some NP like
charged Higgs is at work here. The decay $B^+\to D^0\tau^+\nu$ being 
sensitive to different couplings of $H^\pm$ can contribute significantly 
to this discussion but formfactor uncertainties make this decay less
theoretically clean. A thorough analysis of this decay is presented 
in \cite{Nierste:2008qe} where further references can be found.

Interestingly, the tension between  theory and experiment in the 
case of $Br(B^+\to\tau^+\nu)$
 increases in the presence of a tree level $H^\pm$
exchange which interferes destructively with the $W^\pm$ contribution. 
As addressed long time ago by 
Hou \cite{Hou:1992sy} and in 
modern times calculated first by Akeroyd and Recksiegel 
\cite{Akeroyd:2003zr}, and later by
Isidori and Paradisi \cite{Isidori:2006pk}, 
one has in the MSSM with MFV and large $\tan\beta$
 \be\label{BP-MSSM}
\frac{Br(B^+\to\tau^+\nu)_{\rm MSSM}}
{Br(B^+\to\tau^+\nu)_{\rm SM}}=
\left[1-\frac{m_B^2}{m^2_{H^\pm}}\frac{\tan^2\beta}{1+\epsilon\tan\beta}
\right]^2,
\ee
with $\epsilon$ collecting the dependence on supersymmetric parameters.
This means that in the MSSM this decay can be strongly suppressed unless 
the choice of model parameters is such that the second term in the parenthesis
is larger than 2. 
Such a
possibility that would necessarily imply a light charged Higgs and large $\tan\beta$ values
seems to be very unlikely in view of the constraints from other 
observables~\cite{Antonelli:2008jg}.
Recent summaries of  $H^\pm$ physics can be found in
\cite{Barenboim:2007sk,Ellis:2009qx}.

{\bf Goal 10: Rare Kaon Decays}

Among the top highlights of flavour physics in the next decade
will be the measurements of the branching ratios of two {\it golden modes}
$\kpn$ and $\klpn$. $\kpn$ is CP conserving while $\klpn$ is governed by 
CP violation. Both decays are dominated in the SM and  many of its
extensions by $Z$ penguin contributions.
It is well known that these decays are theoretically 
very clean and are known in the SM including NNLO QCD corrections and 
electroweak corrections \cite{Buras:2005gr,Buras:2006gb,Brod:2008ss}.
Moreover, extensive calculations of isospin breaking effects and 
non-perturbative effects have been done \cite{Isidori:2005xm,Mescia:2007kn}.
The 
present theoretical uncertainties in $Br(\kpn)$ and $Br(\klpn)$ are 
at the level of $2-3\%$ and $1-2\%$, respectively.

We will discuss these decays in more detail in Section 4 but let me stress
already here that the measurements of their branching ratios with an
accuraccy of $10\%$ will give us a very important insight into the physics 
at short distance scales. NA62 at CERN in the case of $\kpn$ and KOTO at 
J-PARC in the case of $\klpn$ will tell us how these two decays are
affected by NP.

The decays $K_L\to\pi^0l^+l^-$ are not as 
theoretically clean as the $K\to\pi\nu\bar\nu$ chanels and are less sensitive 
to NP contributions but they probe different operators beyond the SM and 
having accurate branching ratios for them would certainly be useful. 
Further details on this decay can be found in 
\cite{Buchalla:2003sj,Isidori:2004rb,Friot:2004yr,Mescia:2006jd,Prades:2007ud,Buras:1994qa}.

{\bf Goal 11: Rare B Decays $\mathbf{B\to X_s\nu\bar\nu}$, $\mathbf{B\to K^*\nu\bar\nu}$ and 
        $\mathbf{B\to K\nu\bar\nu}$}

Also  $B$ decays with $\nu\bar\nu$ in the final state provide a very
good test of modified $Z$ penguin contributions 
\cite{Colangelo:1996ay,Buchalla:2000sk}, 
but their measurements appear to be
even harder than those of the rare K decays just discussed. Recent analyses
of these decays within the SM and several NP scenarios can be found in 
\cite{Altmannshofer:2009ma,Bartsch:2009qp}.

The inclusive decay $B\to X_s\nu\bar\nu$ is theoretically as clean as 
$K\to\pi\nu\bar\nu$ decays but the parametric uncertainties are a bit
larger. The two exclusive channels are affected by  formfactor uncertainties 
but recently in the case of $B\to K^*\nu\bar\nu$ \cite{Altmannshofer:2009ma} 
and $B\to K\nu\bar\nu$ 
\cite{Bartsch:2009qp}
significant progress has been made. In the latter paper this has been achieved 
by considering simultaneously also $B\to K l^+l^-$.
Very recently non-perturbative tree level contributions from $B^+\to
\tau^+\nu$ to $B^+\to K^+\nu\bar\nu$ and $B^+\to K^{*+}\nu\bar\nu$ at the 
level of roughly $10\%$ 
have been pointed out \cite{Kamenik:2009kc}.

The interesting feature of these three $b\to s\nu\bar\nu$ transitions, in particular when 
taken together, is their sensitivity to right-handed currents 
\cite{Altmannshofer:2009ma}. Super-B
machines should be able to measure them at a satisfactory level.

{\bf Goal 12: Lattice Calculations of Hadronic Matrix Elements in 
$\mathbf{\epe}$}

One of the important actors of the previous decade in flavour physics was the ratio
$\epe$
 that measures the size of the direct CP
violation in $K_L\to\pi\pi$ 
relative to the indirect CP violation described by $\varepsilon_K$. 
In the SM $\varepsilon^\prime$ is governed by QCD penguins but 
receives also an important distructively interfering
 contribution from electroweak
penguins that is generally much more sensitive to NP than the QCD
penguin contribution.

Here the problem is
the strong cancellation of 
QCD penguin contributions and electroweak penguin contributions to
 $\epe$ and in order to obtain useful predictions  the precision on 
the corresponding hadronic parameters
$B_6$ and $B_8$ should be at least $10\%$. 
Lattice theorists around Norman Christ hope to make progress on $B_6$, 
$B_8$ and other $\epe$ related hadronic matrix elements in the coming
decade. 
This would really be good, as the
calculations of  short distance contributions to this ratio (Wilson 
coefficients of QCD and electroweak penguin operators) have been 
 known already 
for 16 years at the NLO level \cite{Buras:1993dy,Ciuchini:1993vr}
and present technology could extend them 
to the NNLO level if necessary. 

The present experimental world average  from 
NA48 \cite{Batley:2002gn}   and 
KTeV \cite{AlaviHarati:2002ye,Worcester:2009qt}, 
\be
\epe=(16.8\pm 1.4)\cdot 10^{-4}~,
\ee
could have an important impact on several extentions of the SM discussed
in Section 4 if $B_6$ and $B_8$ were known.
An analysis of $\epe$ in the LHT model demonstrates this problem
in explicit terms \cite{Blanke:2007wr}. If one uses $B_6=B_8=1$ as obtained 
in the large N approach, $(\epe)_{\rm SM}$ is in the ballpark of the 
experimental data and sizable departures of $Br(\klpn)$ from its SM 
value are not allowed. $\kpn$ being CP conserving and consequently 
not as strongly correlated with $\epe$ as $\klpn$ could still be 
enhanced by $50\%$. On the other hand if $B_6$ and $B_8$ are different 
from unity and $(\epe)_{\rm SM}$ disagrees with experiment, much more
room for enhancements of rare K decay branching ratios through
NP contributions is available.
Reviews of $\epe$ can be found in 
\cite{Buras:2003zz,Pich:2004ee}.

{\bf Goal 13: CP Violation in Charm Decays, $\mathbf{D^+(D^+_s)\to l^+\nu}$
and $\mathbf{D^0\to \mu^+\mu^-}$}

Charm physics  has been for many years shadowed by the successes of
$K$ decays and $B$ decays, although a number of experimental groups and 
selected theorists have made a considerable
effort to study them. This is due to the GIM mechanism being very effective in
suppressing the FCNC transitions in this sector, long distance contributions
plaguing the evaluation of  $\Delta M_D$ and insensitivity to top
physics in the loops. However, the large $D^0-\bar D^0$ mixing discovered
in 2007 \cite{Aubert:2007wf,Staric:2007dt,Abe:2007rd} 
and good prospects for the study of CP violation in the above decays at
Super Belle and Super $B$ in Frascati gave a new impetus to this field. 
The main targets
here are:
\begin{itemize}
\item
Dedicated studies of CP Violation in $D$ decays that is predicted to be
very small in the SM, but could be strongly enhanced beyond the SM
 and is theoretically much cleaner than $\Delta M_D$,
\item
Dedicated studies of $D^+\to\mu^+\nu_\mu$, $D^+\to\tau^+\nu_\tau$ and
$D_s\to \tau^+\nu_\tau$ with higher experimental and lattice accuracy
with the aim to study charged Higgs effects,
\item
Rare decays $D^0\to \mu^+\mu^-$ and $D_s\to \mu^+\mu^-$~.
\end{itemize}
Excellent reviews can be found in \cite{Bianco:2003vb,Burdman:2003rs}.
Various aspects of charm physics are discussed in 
\cite{Bigi:2000wn,Falk:2004wg,Falk:2001hx,Grossman:2006jg,Nir:2007ac,Blum:2009sk,Grossman:2009mn,Bigi:2009df}.

The first possible sign of NP appeared in $D_s^+\to l^+\nu$ decays some 
time ago and in 2008 a  $3\sigma$ discrepancy between the SM and the 
data has been declared. Meanwhile this tension decreased to $2\sigma$ 
and in order to have a clear picture we have to wait for the new data
with higher statistics and further improved lattice calculations of the
relevant $D$ weak  decay constants even if the latter 
are already rather precise \cite{Kronfeld:2008gu}.
In fact this example shows how much fun we will have to compare the 
data with theory when both experiments and lattice calculations improve.

{\bf Goal 14: CP Violation in the Lepton Sector and
                     $\mathbf{\theta_{13}}$}

The mixing angles $\theta_{12}$  and $\theta_{23}$ are already known 
with respectable precision. The obvious next targets in this field
are $\theta_{13}$ and the CP phase $\delta_{\rm PMNS}$.
Clearly the discovery of CP violation in the lepton sector 
would be a very important mile stone
in particle physics for many reasons. In particular the most efficient
explanations of the BAU these days follow from leptogenesis. While in the past the
necessary size of CP violation was obtained from new sources of CP
violation at very high see-saw scales, the inclusion of flavour effects,
in particular in  resonant leptogenesis, gave hopes for the explanation
of the BAU using only the phases in the PMNS matrix. This implies certain
conditions for the parameters of this matrix, that is the relevant
$\delta_{\rm PMNS}$, two Majorana phases and $\theta_{13}$. As there was 
a separate talk on neutrino physics at this conference let me just refer 
to this talk and  the review in 
\cite{GonzalezGarcia:2007ib} for the relevant references. 
A recent review 
of models for neutrino masses is given in \cite{Altarelli:2009km}.

{\bf Goal 15: Tests of $\mathbf{\mu-e}$ and $\mathbf{\mu-\tau}$ Universalties}

Lepton flavour violation (LFV) and the related breakdown of universality 
 can be tested in meson decays by studying the
ratios \cite{Masiero:2005wr,Masiero:2008cb}
 \be\label{BP-munu}
R_{\mu e}=\frac{Br(K^+\to\mu^+\nu)}
{Br(K^+\to e^+\nu)}, \qquad
R_{\mu\tau}=
\frac{Br(B^+\to\mu^+\nu)}
{Br(B^+\to \tau^+\nu)},
\ee
where the sum over different neutrino flavours
is understood.
The first case is a high precision affair both for experimentalists and 
theorists as both groups decreased the uncertainties in $R_{\mu e}$ 
well below $1\%$  with a precision of $0.5\%$ recently achieved at
CERN. It will continue to constitute an important test of the 
$\mu-e$ universality.
The ratio $R_{\mu\tau}$ is even more sensitive to NP contributions but
it will still take some time before it will be known with good precision.

{\bf Goal: 16 Flavour Violation in Charged Lepton Decays }

 The search for  LFV clearly belongs to the most important goals in
 flavour physics.
  The non-vanishing neutrino masses and neutrino oscillations as well as
 the see-saw mechanism for the generation of  neutrino masses have 
 given an
 impressive impetus to the study of flavour violation in the lepton
 sector in the last ten years. In the SM with  
right-handed Dirac neutrinos, the smallness 
 of neutrino masses implies tiny branching ratios for LFV processes.
 For instance
\be
Br(\mu\to e\gamma)_{\rm SM}\approx 10^{-54},
\ee
which is more than 40 orders of magnitude below the $90\%$ C.L. upper bound
from the MEGA Collaboration \cite{Brooks:1999pu}
\be\label{ueg}
Br(\mu\to e\gamma)< 1.2 \cdot 10^{-11}.
\ee
Therefore any observation of LFV would be a clear sign of NP.
While we hope that new flavoured leptons will be observed at the LHC, 
even if this will not turn out to be the case, LFV has the following virtue:
sensitivity to short distance scales as high as $10^{10}-10^{14}\gev$, in 
particular when the see-saw mechanism is at work.

The prospects
for the measurements of LFV processes with much higher sensitivity than
presently available in the next decade look very good. 
In particular the MEG experiment at PSI \cite{Yamada:2005tg}
should be able to test $Br(\mu\to e\gamma)$ at the level of
$\ord(10^{-13}-10^{-14})$, and the Super Flavour Factory \cite{Bona:2007qt} is
planned to reach a sensitivity for $Br(\tau\to\mu\gamma)$ of at least
$\ord(10^{-9})$. The planned accuracy of SuperKEKB of $\ord(10^{-8})$ for
  $\tau\to\mu\gamma$ is also of great interest. Very important will also be an
improved upper bound on $\mu-e$ conversion in Ti. In this context the
dedicated J-PARC experiment PRISM/PRIME \cite{Mori} should reach the
sensitivity of $\ord(10^{-18})$. This means an improvement by six orders of magnitude relative to the present upper bound from SINDRUM II at PSI \cite{Dohmen:1993mp}.

Now the various supersymmetric models, the LHT model and the RS models 
are capable of reaching the
bound in (\ref{ueg}) and in fact this bound puts already rather
stringent constraints on the parameters of these models. For instance in
the case of the LHT model the mixing matrix in the mirror lepton sector
has to be either very hierarchical, at least as hierarchical as the CKM
matrix or the mirror-lepton spectrum has to be quasi-degenerate 
\cite{Blanke:2007db,delAguila:2008zu,Blanke:2009am}. Analogous
constraints exist in other models. We will discuss some aspects of LFV in 
Section 4.

In order to distinguish various NP scenarios that come close to the
bound in (\ref{ueg}) it will be essential to study a large set
of decays to three leptons in the  final state. Indeed, while in the
MSSM \cite{Ellis:2002fe,Arganda:2005ji,Brignole:2004ah,Paradisi:2005tk,Paradisi:2006jp}
the dominant role in the decays with three leptons in the final
state and in $\mu-e$ conversion in nuclei is played by the dipole
operator, in \cite{Blanke:2007db,delAguila:2008zu} it was found that this operator is 
much less relevant in
the LHT model, with $Z^0$ penguin and box diagrams being the dominant
contributions. This implies a striking difference between various ratios of
branching ratios of type $Br(l_i\to 3l_j)/Br(l_i\to l_j\gamma)$ in the
MSSM, where they are typically $\ord(10^{-2}-10^{-3})$ and in the LHT 
model, where they are $\ord(10^{-1})$ \cite{Blanke:2009am}. A very recent
short 
review of these topics can be found in \cite{Feldmann:2009tp}.

There exist also interesting correlations between leptogenesis and
LFV but this is beyond the scope of this
presentation. Additional correlations relevant for LFV
 will be discussed in Section 4.

{\bf Goal 17: Electric Dipole Moments}

CP violation has only been observed in flavour 
violating processes. Non-vanishing electric dipole moments  signal 
CP violation in  flavour conserving transitions. 
In the SM CP violation in flavour conserving
processes is very strongly suppressed as best expressed by 
the SM values of electric dipole moments of the neutron and electron
that amount to \cite{Pospelov:2005pr}
\begin{equation}
d_n\approx 10^{-32}~ {\rm e~ cm},\qquad d_e\approx 10^{-38}~ {\rm e~ cm.}
\end{equation}  

This should be compared with the present experimental bounds 
\cite{Baker:2006ts,Regan:2002ta}
\begin{equation}
d_n \le 2.9\cdot  10^{-26}~ {\rm e~ cm.}\qquad 
d_e\le 1.6\cdot10^{-27}~ {\rm e~ cm.}
\end{equation}  
They should be improved in the coming years by 1-2 orders of magnitude.

Similar to LFV, an observation of a non-vanishing EDM  would 
imply necessarily NP at work. Consequently  correlations between LFV 
and EDM's in specific NP scenarios are to be expected, in particular in 
supersymmetric models, as both classes of observables 
are governed by dipole operators. A rather complete analysis of such
correlations has been recently presented in \cite{Hisano:2008hn} where
further references can be found. 
We will encounter some specific examples in Section 4.

{\bf Goal 18: Clarification of the $\mathbf{(g-2)_\mu}$ Anomaly}

The measured anomalous magnetic moment of the electron, $(g-2)_e$, is in 
an excellent agreement with SM expectations. On the other hand, the measured 
 anomalous magnetic moment of the muon, $(g-2)_\mu$, is significantly larger 
than its SM value.
The most recent SM prediction reads \cite{Prades:2009qp} 
\be\label{aSM}
a_\mu^{\rm SM}=11659~1834~(49)\cdot 10^{-11}
\ee
and the experimental value from BNL \cite{Bennett:2002jb,Bennett:2004pv}
\be\label{aEXP}
a_\mu^{\rm exp}=11659~2080~(63)\cdot 10^{-11},
\ee
where $a_{\mu}\!=\!(g-2)_{\mu}/2$. Consequently, 
\be
\Delta a_{\mu}\!=\!a_{\mu}^{\rm exp}\!-\!a_{\mu}^{\rm SM}= (2.5\pm0.8)
\times 10^{-9}~,
\ee
implying a $3.1\sigma$ deviation from the SM value. Similar results can be 
found in \cite{Passera:2004bj,Passera:2005mx}.

Hadronic contributions to $(g-2)_\mu$ make the comparison of data and theory 
a bit problematic. 
Yet, as this anomaly has been 
with us already for a decade and tremendous effort by a number of theorists
has been made to clarify this issue, this anomaly could indeed come from 
NP. 

The MSSM with large $\tan\beta$ and sleptons  with masses below $400\gev$ 
is capable of reproducing the experimental value of $a_\mu$ 
 provided the $\mu$ parameter
in the Higgs Lagrangian has a specific sign, positive in my conventions:
\be
\frac{a^{\rm MSSM}_\mu}{ 1 \times 10^{-9}}
\approx 1.5\left(\frac{\tan\beta }{10} \right)
\left( \frac{300~\rm GeV}{m_{\tilde \ell}}\right)^2 \text{sgn}\,\mu\,.
\label{eq:g_2}
\ee
Moreover an interesting correlation between the amount of necessary
shift $\Delta a_\mu$ and the value of $Br(\tau\to\mu\gamma)$ and 
$Br(\mu\to e\gamma)$ exists \cite{Isidori:2007jw}, 
 implying that these two branching ratios could
be as high as $4\cdot 10^{-9}$ and $3\cdot 10^{-12}$, respectively
and thus in the reach of dedicated experiments in the coming years.
Other correlations of this type in supersymmetric flavour models 
will be discussed in Section 4.
On the other hand the LHT fails to reproduce the data in (\ref{aEXP}) and $a_\mu$ in this
model is within the uncertainties indistiguishable from its SM value
\cite{Choudhury:2006sq,Blanke:2007db}.
Apparently there is no visible correlation between NP in $a_\mu$ and LFV 
in this model.
Thus if the data in (\ref{aEXP}) remain, they would favour the MSSM over the 
LHT.
Recent reviews on $(g-2)_\mu$ can be found in 
\cite{Passera:2008jk,Passera:2008hj,DeRafael:2008iu,Jegerlehner:2009ry,Prades:2009qp}.

{\bf Goal 19: Flavour Violation at High Energy}

Our presentation deals mainly with tests of flavour and CP violation in
low energy processes. However, at the LHC it will be possible to investigate
 these phenomena also in high energy processes, in particular in top quark
decays.
Selected recent analyses on flavour physics in high energy processes
can be found in \cite{Grossman:2007bd,Feng:2007ke,Agashe:2007ki,Agashe:2006hk,Agashe:2006wa,Agashe:2008jb,Hiller:2008wp,Drobnak:2008br}.

{\bf Goal 20: Construction of a New Standard Model (NSM)}

Finally, in view of so many parameters present in basically all extensions
of the SM like the MSSM, the LHT model and RS models, it is unlikely from my point of view 
that any of the models
studied presently in the literature will turn out in 2026 to be the new
model of elementary particle physics. On the other hand various structures, 
concepts
and ideas explored these days in the context of specific models may well turn
out to be included in the NSM that is predictive, consistent with all the 
data and giving explanation of observed hierarchies in fermion masses and
mixing matrices. While these statements may appear to be very naive, it is a
fact that the construction of the NSM is the main goal of elementary particle 
physics
and every theorist, even as old as I am,
 has a dream that the future NSM will carry her (his) name.

\section{Waiting for Signals of New Physics in FCNC Processes}

\begin{figure}[htbp]
\begin{center}
\includegraphics[width=2.7in]{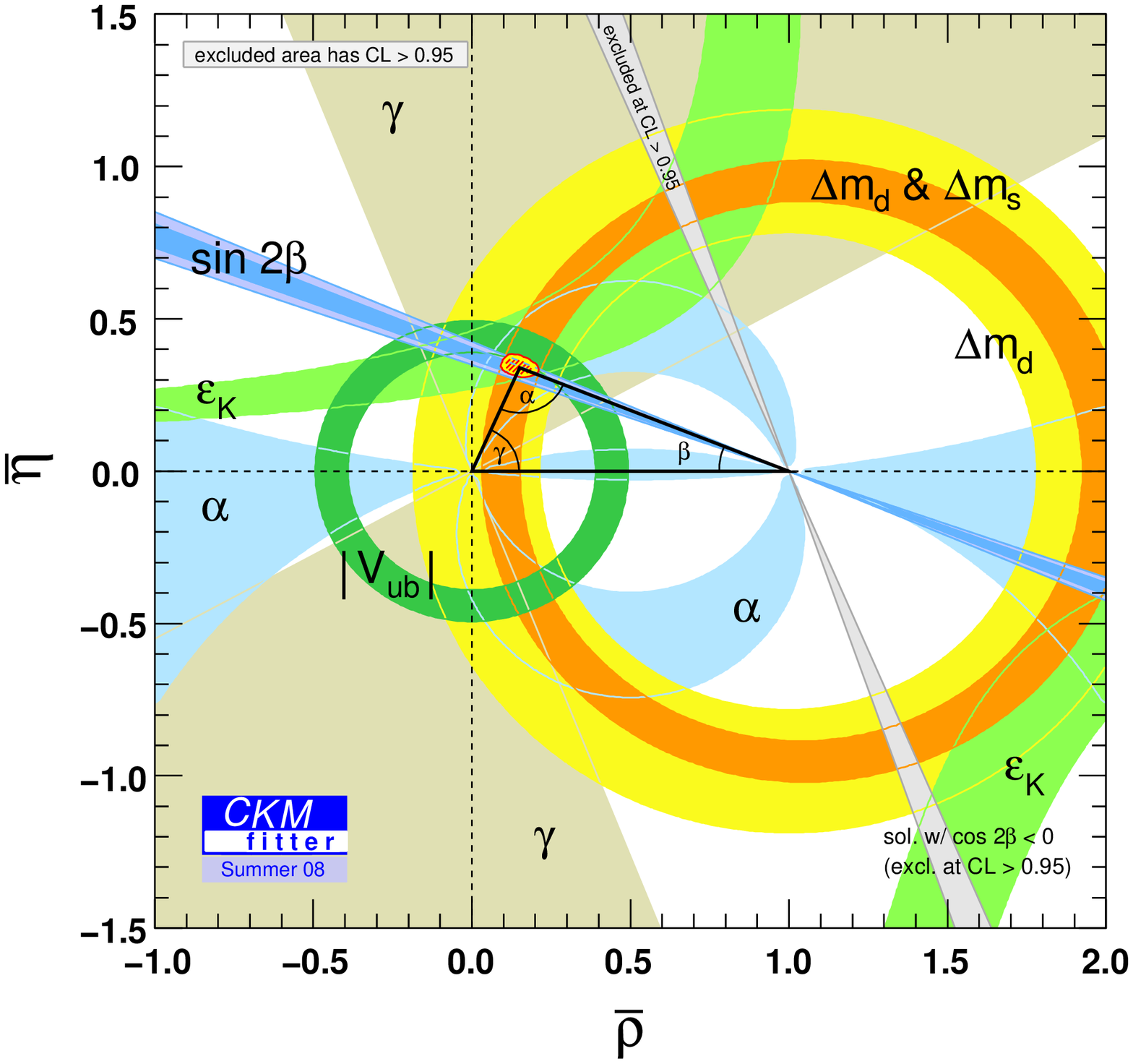}
\includegraphics[width=2.7in]{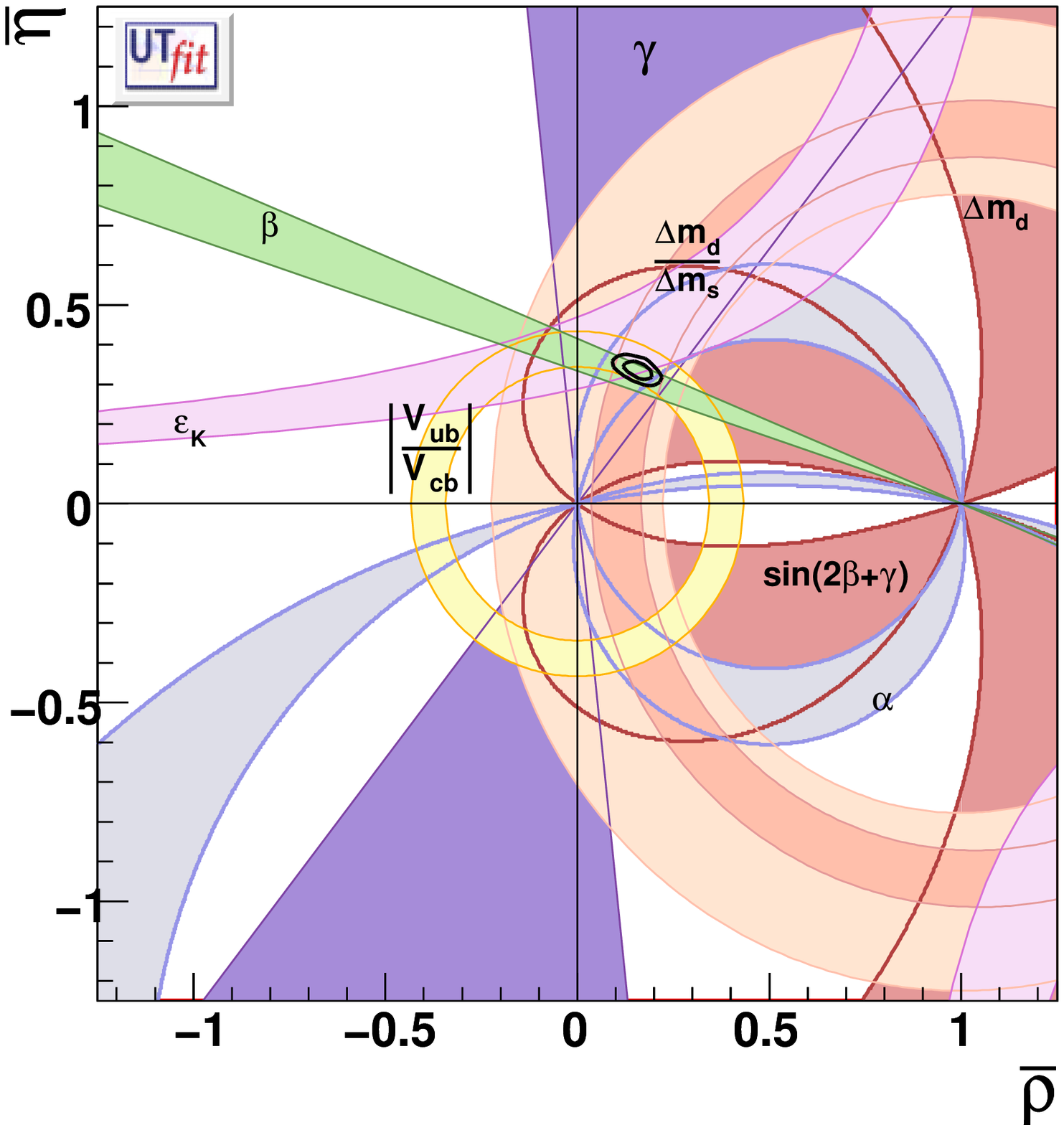}
\end{center}
\caption{\label{UT12} Unitarity triangle fits by the
CKMfitter \cite{Charles:2004jd} (left) and 
UTfit \cite{Bona:2006sa} (right) 
collaborations in 2009.}
\end{figure}

\subsection{A Quick Look at the Status of the CKM Matrix}
The success of the CKM description of flavour violation and in particular CP 
violation can be best seen by looking at the so-called 
Unitarity Triangle (UT) fits in Fig.\ref{UT12}. 
The extensive analyses of the UTfit and CKMfitter collaborations 
\cite{UTfit,CKMfitter} show that the data 
on  $\vus,~\vub,~\vcb,~\varepsilon_K$, $\Delta M_d$, $\Delta M_s$ and
 the CP-asymmetry
$S_{\psi K_S}$, that measures the angle $\beta$ in the UT,
are compatible with each other within theoretical and experimental 
uncertainties. Moreover the angles $\alpha$ and $\gamma$ of the UT
 determined by means of various non-leptonic decays and sophisticated
 strategies are compatible with the ones extracted from Fig.~\ref{UT12}.

While this agreement is at first sight impressive and many things could 
already have turned out to be wrong, but they did not, one should remember 
that only very few theoretically clean observables have been measured
precisely so far.
 
 The three 
parameters relevant for the CKM matrix that have been measured accurately
are:
\be\label{CKMpar}
\vus=0.2255\pm0.0010, \qquad \vcb=(41.2\pm1.1)\cdot 10^{-3}, \qquad
\beta=\beta_{\psi K_S}=(21.1\pm0.9)^\circ,
\ee
where the last number follows from \cite{Barberio:2007cr}
\be\label{s2beta}
\sin 2\beta=0.672\pm0.023.
\ee

It should be mentioned that the value for $\vcb$ quoted above results
from inclusive and exclusive decays that are not fully consistent with
each other. Typically the values resulting from exclusive decays are 
below $40\cdot 10^{-3}$. As the value of $\vcb$ is very important for FCNC
processes in the $K$ system it would be important to clarify this difference
which has been with us already for many years. Hopefully, the future Super B
facilities in Italy and Japan and new theoretical ideas 
will provide more precise values. More on this can be found in Bevan's talk
\cite{Bevan:2009qe}.

\subsection{Strategies in the Present Decade}
\begin{figure}
\begin{center}
\includegraphics[width=3in]{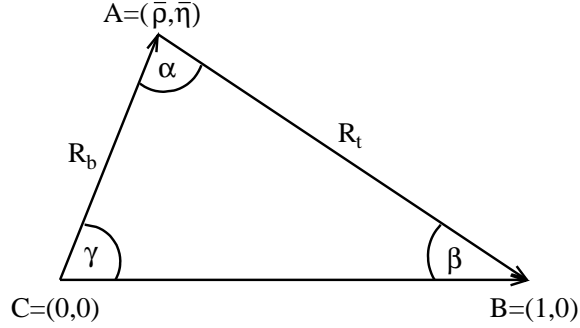}
\end{center}
\caption{\label{fig:utriangle}The Unitarity Triangle}
\end{figure}

The strategies for the determination of the UT in this decade used basically  the following
set of fundamental variables:
\begin{equation}\label{eq:UT_parameter_2}
\vus \equiv \lambda ~,~~~ \vcb ~,~~~ R_t ~,~~~ \beta~,
\end{equation}
where (see Fig.~\ref{fig:utriangle})
\begin{equation}
R_t \equiv \frac{| V_{td}^{}V^*_{tb}|}{| V_{cd}^{}V^*_{cb}|} 
= \sqrt{(1-\bar\varrho)^2 +\bar\eta^2}
= \frac{1}{\lambda} \left| \frac{V_{td}}{V_{cb}} \right| ~,\qquad
V_{td} = |V_{td}| e^{- i \beta} .
\end{equation}

Now, $\vus$ and $\vcb$ extracted from tree level decays are free from NP
pollution.
In contrast, $R_t$ and $\beta$ in the parameter set in 
(\ref{eq:UT_parameter_2}) can only be extracted from loop-induced 
FCNC processes and hence are potentially sensitive to NP effects. 
Consequently, the corresponding UT, the universal unitarity triangle 
(UUT)~\cite{Buras:2000dm} of models with CMFV \cite{Buras:2003jf,Blanke:2006ig}, could differ from the 
true UT triangle.

Indeed,
within the SM and CMFV models the parameters $R_t$ and $\beta$ can be related in the
following way to the observables $\Delta M_{s,d}$ and $S_{\psi K_S}$
\begin{equation} \label{eq:Rt_sin2beta_SM}
R_t = \xi \frac{1}{\lambda} \sqrt{\frac{m_{B_s}}{m_{B_d}}} \sqrt{\frac{\Delta M_d}{\Delta M_s}} ~,~~~\quad \sin2\beta = S_{\psi K_S}~,
\end{equation}
where $\Delta M_d$ and $\Delta M_s$ are the mass differences in 
the neutral $B_d$ and $B_s$ systems,
$S_{\psi K_s}$ represents the mixing-induced CP asymmetry in the decay $B_d \to \psi K_S$
and the value of the non-perturbative parameter $\xi$ is given 
as follows
\be\label{xi}
\xi = 
\frac{\sqrt{\hat B_{B_s}}F_{B_s} }{ \sqrt{\hat B_{B_d}}F_{B_d}}=1.21\pm 0.04,
\qquad \xi=1.258\pm0.033
\ee
as summarized  by Lubicz and Tarantino \cite{Lubicz:2008am} and by  
the HPQCD collaboration 
\cite{Gamiz:2009ku}, 
respectively.

In the presence of NP however, these relations are modified and one finds
\begin{equation} \label{eq:Rt_sin2beta_NP}
R_t = \xi \frac{1}{\lambda} \sqrt{\frac{m_{B_s}}{m_{B_d}}} \sqrt{\frac{\Delta M_d}{\Delta M_s}} \sqrt{\frac{C_{B_s}}{C_{B_d}}} ~,~~~\quad  \sin(2 \beta + 2 \phi_{B_d}) = S_{\psi K_s} ~,
\end{equation}
where the NP phases $\phi_{B_q}$  in $B_q$ 
mixing and the parameters $C_q$ ($q=d,s$) are defined through
\cite{Bona:2005eu}
\begin{equation} \label{eq:M12d}
M_{12}^q = \langle B_q| H_{\rm eff}^q |\bar B_q \rangle = (M_{12}^q)^{\rm
  SM}+ (M_{12}^q)^{\rm NP} =  C_{B_q}e^{2 i\phi_{B_q}}(M_{12}^q)^{\rm SM}. 
\end{equation}
For the mass differences in the $B_q$ meson system one then has
\begin{equation}
\Delta M_q = 2 |M_{12}^q| =  C_{B_q} \Delta M_q^{\rm SM} ~.
\end{equation}

The outcome of using (\ref{eq:Rt_sin2beta_SM}), $\vcb$ and $\vus$ are
the parameters $\bar\varrho$ and $\bar\eta$ that presently 
are given as follows
\begin{displaymath}\label{rhobar}
 \bar\varrho=\left\{
\begin{array}{ll}
0.154\pm0.021 \,  & \mbox{(UTfit),}\\
0.139^{+0.025}_{-0.027} \,  & \mbox{(CKMfitter).}
\end{array}
\right.
\end{displaymath}

\begin{displaymath}\label{etabar}
 \bar\eta=\left\{
\begin{array}{ll}
0.340\pm0.013 \,  & \mbox{(UTfit),}\\
0.341^{+0.016}_{-0.015} \,  & \mbox{(CKMfitter).}
\end{array}
\right.
\end{displaymath}
 Yet, this determination could be polluted by NP and as we will see below 
another look at the UT analysis presented below reveals a number of 
tensions in this determination.

Finally let us stress that the angle $\alpha$ is already well determined from 
$B_d\to\varrho\varrho$ and $B_d\to\varrho\pi$ decays 
\cite{Barberio:2007cr}:
\be
\alpha=(91.4\pm 4.6)^\circ.
\ee
A specific analysis employing the mixing induced CP asymmetries 
$S_{\psi K_S}$, $S_{\varrho\varrho}$
 and the QCDF approach finds \cite{Bartsch:2008ps} $\alpha=(87\pm6)^\circ$. 
Summaries of other determinations of $\alpha$ exist
\cite{Buchalla:2008jp,Antonelli:2009ws}.

\subsection{Unitarity Triangle in the LO Approximation}
Even if NP could have still some visible impact on the determination of
the UT presented above, the basic shape of the UT has been determined 
in this decade and in the LO approximation it can be characterized by
two numbers:
\be
\alpha=90^\circ, \qquad \sin 2\beta =\frac{2}{3},
\ee
implying rather accurately
\be
\beta=21^\circ, \qquad \gamma= 69^\circ,
\ee

\be 
\bar\varrho=\sin\beta\cos\gamma=0.128, \qquad \bar\eta=\sin\beta\sin\gamma=0.33
\ee
and
\be
R_b=\sin\beta=0.36, \qquad R_t=\sin(\alpha+\beta)=0.93~.
\ee

This is an important achievement of the present decade but in my opinion 
in the next decade we should proceed in a different manner. First, however
let us briefly return to our first goal of the previous section.

 \subsection{The Quest for $\mathbf{\vub}$ and the Angle $\mathbf{\gamma}$}
 As we have already stressed in  Goal 1 of the previous section,
 precise measurements of the side $R_b(\vub)$ and of the angle $\gamma$ 
in the UT of 
 Fig.~\ref{fig:utriangle} by means of
 tree level decays  that are independent of any new physics to a good
 approximation, are undisputably very important.

Indeed
the status of $|V_{ub}|$ and $\gamma$ from tree level decays
is not  particularly impressive:
\begin{displaymath}\label{vub}
|V_{ub}| =\left\{
\begin{array}{ll}
(4.0\pm0.3)\cdot 10^{-3} \,  & \mbox{(inclusive),}\\
(3.5\pm0.4)\cdot 10^{-3} \,  & \mbox{(exclusive),}
\end{array}
\right.
\end{displaymath}
\begin{displaymath}\label{gamma}
 \gamma=\left\{
\begin{array}{ll}
(78\pm12)^\circ \,  & \mbox{(UTfit),}\\
(76^{+16}_{-23})^\circ \,  & \mbox{(CKMfitter).}
\end{array}
\right.
\end{displaymath}

It is very important to precisely measure  $|V_{ub}|$ and $\gamma$ in tree 
level decays in
the future as they determine the so-called reference UT (RUT) 
\cite{Goto:1995hj}, 
that is free from NP
pollution. 
Having determined $\vub$ and $\gamma$ from tree level decays  would allow to 
obtain the CKM matrix without NP pollution,
 with the four 
fundamental flavour parameters being now
\be
\vus,\qquad \vub,\qquad \vcb,\qquad \gamma~,
\ee
and to construct the RUT \cite{Goto:1995hj} 
by means of 
$\lambda=\vus$,
\be
R_b
= \left(1-\frac{\lambda^2}{2}\right)\frac{1}{\lambda}
\left| \frac{V_{ub}}{V_{cb}} \right|,
\label{2.95}
\ee  
and $\gamma$.

This is indeed a very important goal as it would give us immediately 
the true values of $R_t$ and $\beta$ 
 in Fig.~\ref{fig:utriangle} by simply using
\be\label{RTBETAG}
R_t=\sqrt{1+R_b^2-2 R_b\cos\gamma},\qquad
\cot\beta=\frac{1-R_b\cos\gamma}{R_b\sin\gamma}.
\ee
Comparing this result with the one obtained by means of 
(\ref{eq:Rt_sin2beta_SM}) and using 
(\ref{eq:Rt_sin2beta_NP}) would tell us whether NP in $\Delta B=2$ processes is at work.

\subsection{Strategies for the Search for New Physics in the Next Decade}
Let us first emphasize that until now only $\Delta F=2$ FCNC processes 
 could be used in the UTfits. The measured $B\to X_s\gamma$ and $B\to X_s
 l^+l^-$ decays  and their exclusive 
counterparts are sensitive to $\vts$ that has nothing to do with the plots in 
Fig.~\ref{UT12}. The same applies to the observables in the $B_s$ system, which
with the $S_{\psi\phi}$ anomaly observed by CDF and D0 and the studies 
of rare $B_s$ decays at Tevatron and later at  LHC are becoming central
for flavour physics. Obviously these comments also apply to all lepton 
flavour violating processes.

In this context 
a special role is played by $Br(\kpn)$ and $Br(\klpn)$ as their values
allow a theoretically clean construction of the UT in a manner complementary 
to its present determinations \cite{Buchalla:1994tr}: the height of the UT 
is determined from $Br(\klpn)$ and the side $R_t$ from $Br(\kpn)$. Thus 
projecting the results of future experimental results for these two 
branching ratios on the $(\bar\varrho,\bar\eta)$ plane could 
be a very good test of the SM.

Yet, generally 
I do not think that in the 
context of the search for the NSM (see Goal 20) it is a good strategy  to
project the results of all future measurements of rare decays 
on the $(\bar\varrho,\bar\eta)$ plane or any other of five planes 
related to the remaining unitarity triangles. This would only teach us about
possible inconsistences within the SM but would not point towards 
a particular NP model.

In view of this, here comes a proposal for the strategy for searching 
for NP in the next decade, in which hopefully $R_b$ and $\gamma$ will
be precisely measured, CP violation in the $B_s$ system explored and 
many goals listed in the previous section reached.

This strategy, which is a summary of many ideas present already in the
literature, proceeds in three steps:

{\bf Step 1}

In order to study transparently possible tensions between 
$\varepsilon_K$, $\sin 2\beta$, $\vub$, $\gamma$ and $R_t$ let us 
leave the $(\bar\varrho,\bar\eta)$ plane and {go to}
the $R_b-\gamma$ plane \cite{Buras:2002yj} suggested 
already several years ago and recently strongly supported by the 
analyses in 
\cite{Altmannshofer:2007cs,Altmannshofer:2009ne}. 
The $R_b-\gamma$ plane is shown in Fig.~\ref{fig:UTfit}. We will explain this 
figure in the next subsection.

{\bf Step 2}

In order to search for NP in rare $K$, $B_d$, $B_s$, $D$ decays, in 
CP violation in $B_s$ and charm decays, in LFV decays, in EDM's and 
$(g-2)_\mu$ let us go to specific plots that exhibit correlations 
between various observables. As we will see below such correlations 
will be crucial to 
distinguish various NP scenarios.
Of particular importance are the correlations between the 
CP asymmetry $S_{\psi\phi}$ and
$B_s\rightarrow\mu^+\mu^-$, between 
the anomalies in $S_{\phi K_s}$ and $S_{\psi\phi}$, 
between $\kpn$ and $\klpn$, between $\kpn$ and $S_{\psi\phi}$,
between $S_{\phi K_s}$ and $d_e$, between $S_{\psi\phi}$ and $(g-2)_{\mu}$ and
also those involving lepton flavour violating decays.

{\bf Step 3}

In order to monitor the progress made in the next decade when additional 
data on flavour changing processes will become available, it is useful to
construct a ``DNA-Flavour Test'' of NP models \cite{Altmannshofer:2009ne} 
including Supersymmetry, the LHT
model, the RSc and various supersymmetric flavour models and other models,
with the aim to distinguish between these NP scenarios in a global manner.

Having this strategy in mind we will in the rest of this writing illustrate
it on several examples.

\subsection{The $\mathbf{\varepsilon_K}$-Anomaly and Related Tensions} 
The CP-violating parameter $\varepsilon_K$ in the SM is given as follows
\be\label{epsSM}
|\varepsilon_K|^{\rm SM}=\kappa_\varepsilon C_\varepsilon\hat B_K\vcb^2\vus^2
\left(\frac{1}{2}\vcb^2R_t^2\sin 2\beta\eta_{tt}S_0(x_t)+
R_t\sin\beta(\eta_{ct}S_0(x_c,x_t)-\eta_{cc}x_c)\right),
\ee
where $C_\varepsilon$ is a numerical constant and the SM loop functions
$S_0$ depend on $x_i= m_i^2(m_i)/M_W^2$. The factors $\eta_{ij}$ are
QCD corrections known at the NLO level 
\cite{Buras:1990fn,Herrlich:1993yv,Herrlich:1995hh,Herrlich:1996vf},
 $\hat B_K$ is a non-perturbative parameter and 
$\kappa_\varepsilon$ is explained below.

 Until the discovery of CP violation in the $B_d$ system, 
$\varepsilon_K$ played the crucial role in  tests 
of CP violation, but after the precise measurements of $\sin 2\beta$ and of the
ratio $\Delta M_d/\Delta M_s$ its role in the CKM fits declined because of the
large error in $\hat B_K$. Also for this reason the size of
CP violation in the $K$ and $B$ systems were commonly declared to be 
compatible with each other within the SM. 

This situation changed in 2008 due to
the improved value of $\hat B_K$, the improved determinations of the elements
of the CKM matrix and in particular due to the inclusion of additional
corrections to $\epsilon_K$~\cite{Buras:2008nn} that were neglected in the
past, 
enhancing the role of this CP-violating parameter in the search for NP.

Indeed it has been recently stressed \cite{Buras:2008nn} that 
the SM prediction for $\epsilon_K$ implied by
the measured value of $\sin2\beta$ may be too small to agree with experiment. 
The main reasons for
this are on the one hand a decreased value of 
$\hat B_K=0.724\pm 0.008\pm 0.028$~\cite{Aubin:2009jh} 
(see also \cite{Antonio:2007pb}), lower by 5--10\% with respect to 
the values used in UT fits until recently  
and on the other hand the decreased value of $\epsilon_K$ in the SM 
arising from a multiplicative factor, 
estimated as $\kappa_\epsilon=0.92\pm 0.02$~\cite{Buras:2008nn,Buras:2009pj}. 
Earlier discussions of such corrections can be found in 
\cite{Anikeev:2001rk,Andriyash:2003ym,Andriyash:2005ax}.

\begin{figure}[t]
\includegraphics[width=1.\textwidth]{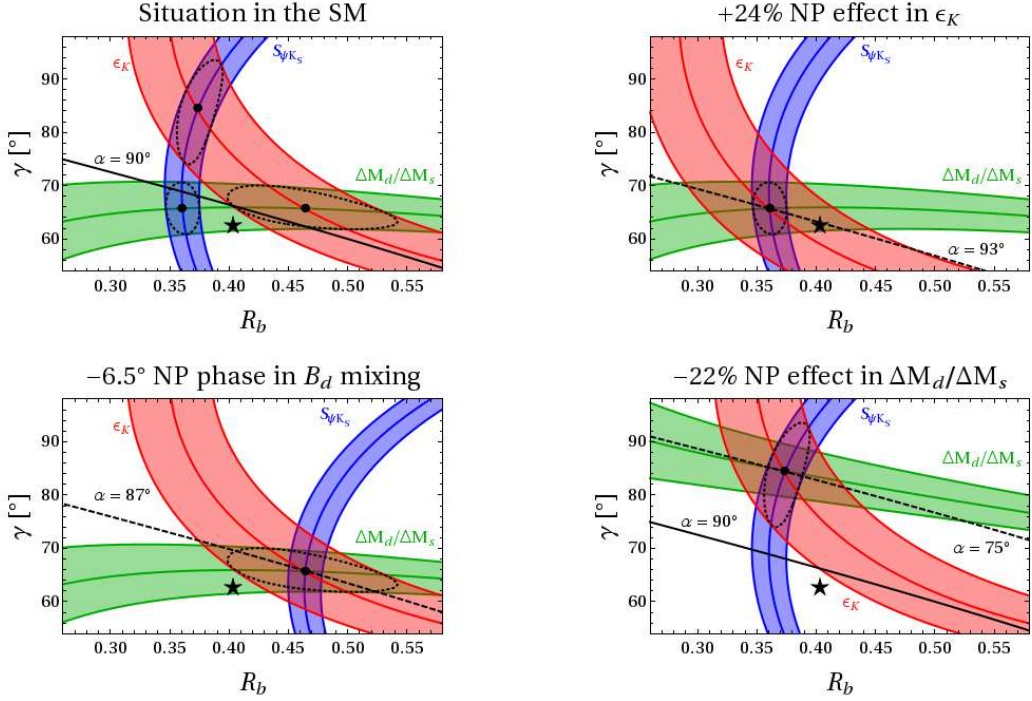}
\caption{\small
The $R_b-\gamma$ plane as discussed in the text. For further explanations see
\cite{Altmannshofer:2009ne}}
\label{fig:UTfit}
\end{figure}

Given that $\epsilon_K \propto\hat B_K \kappa_\epsilon$, the total 
suppression of $\epsilon_K$ 
compared to the commonly used formulae is typically of order $15-20\%$
and using (\ref{CKMpar}), (\ref{s2beta}) and 
(\ref{eq:Rt_sin2beta_SM}) one finds now~\cite{Buras:2009pj}
\begin{equation} \label{eq:epsK_SMnumber}
|\epsilon_K|^{\rm SM} = (1.78 \pm 0.25) \times 10^{-3}~,
\end{equation}
to be compared to the experimental measurement \cite{Amsler:2008zzb}
\begin{equation} \label{eq:epsK_exp}
|\epsilon_K|^{\rm exp} = (2.229 \pm 0.010) \times 10^{-3}~.
\end{equation}
The 15\% error in~(\ref{eq:epsK_SMnumber}) arises from the three main sources of uncertainty that
are still $\hat B_K$, $|V_{cb}|^4$ and $R_t^2$. However, it should be stressed
that $\hat B_K$ known by now with $4\%$ accuracy is not the main uncertainty 
which now is dominantly due to $\vcb$ and in the ballpark of $10\%$.

As seen in (\ref{epsSM})
the agreement between the SM  and (\ref{eq:epsK_exp}) improves for 
higher values of $\hat B_K$, $R_t$ or $|V_{cb}|$ and also 
the correlation between $\epsilon_K$ and $\sin 2 \beta$ within the SM 
is highly sensitive to these parameters. 
Consequently improved determinations of all these parameters 
are very desirable in order to find out whether NP is at work 
in $S_{\psi K_S}$ or in 
$\epsilon_K$ or both. Some ideas can be found
in~\cite{Lunghi:2008aa,Buras:2008nn,Buras:2009pj,Lunghi:2009sm}. 

The tension in question can be also seen in the most recent fit of the
UTfit collaboration shown in Fig.~\ref{UT12}, which now also includes 
the $\kappa_{\varepsilon}$ correction.
In order to see this more transparently 
let us have now a look at the $R_b-\gamma$ plane in 
Fig.~\ref{fig:UTfit} taken from \cite{Altmannshofer:2009ne}, 
where details on input parameters 
can be found. There, in the upper left plot the {\it blue} ({\it green})
region corresponds to the 1$\sigma$ allowed range for $\sin2\beta$ ($R_t$) as
calculated by means of~(\ref{eq:Rt_sin2beta_SM}). The {\it red} region
corresponds to $|\epsilon_K|$ as obtained by equating  (\ref{epsSM}) 
with (\ref{eq:epsK_exp}). Finally the solid black line corresponds to $\alpha=90^\circ$ that is close to the value favoured by UT fits and the determination from $B\to\rho\rho$ \cite{Bartsch:2008ps}.

It is evident that there is a tension between various regions as there are
three different values of $(R_b, \gamma)$, dependending on 
which two constraints
are simultaneously applied. The four immediate solutions to this tension
are as follows:

{\bf 1.}  There is a positive NP effect in $\epsilon_K$ 
while $\sin 2\beta$ and $\Delta M_d/\Delta M_s$ are SM-like~
\cite{Buras:2008nn}, as shown by the upper
right plot of Fig.~\ref{fig:UTfit}.
The required effect in $\epsilon_K$ could be for instance achieved within
models with CMFV by a positive shift in the function
$S_0(x_t)$~\cite{Buras:2009pj} which, while not modifying $(\sin2\beta)_{\psi
  K_S}$ and $\Delta M_d / \Delta M_s$, would require the preferred values of
$\sqrt{\hat B_{B_{d,s}}} F_{B_{d,s}}$ to be by $\simeq 10\%$ lower than the present
central values in order to fit $\Delta M_d$ and $\Delta M_s$
separately. Alternatively, new non-minimal sources of flavour violation
relevant only for the $K$ system could solve the problem.
Note that this solution corresponds to $\gamma \simeq 66^\circ$, 
$R_b \simeq 0.36$ and  $\alpha \simeq 93^\circ$ in accordance with the usual UT analysis.

{\bf 2.} $\epsilon_K$ and $\Delta M_d/\Delta M_s$ are  NP free while $S_{\psi
  K_S}$ is affected by a NP phase $\phi_{B_d}$ in 
$B_d$ mixing of approximately 
$-7^\circ$ as indicated in (\ref{eq:Rt_sin2beta_NP}) and shown by the lower left plot of Fig.~\ref{fig:UTfit}.
The predicted value for $\sin2\beta$ is now shifted to 
$\sin 2\beta\approx 0.85$ 
\cite{Lunghi:2008aa,Buras:2008nn,Lunghi:2009sm,Buras:2009pj}. This value is
  significantly larger than the measured $S_{\psi K_S}$ which allows to fit
  the experimental value of $\epsilon_K$.
Note that this solution is characterized by a large value of $R_b \simeq 0.47$, that is significantly larger than its exclusive determinations but still compatible with the inclusive determinations. The angles $\gamma \simeq 66^\circ$ and $\alpha \simeq 87^\circ$ agree with the usual UT analysis.

{\bf 3. } $\epsilon_K$ and $S_{\psi K_S}$ are NP free while the determination
of $R_t$ through $\Delta M_d/\Delta M_s$ is affected by NP as 
indicated in (\ref{eq:Rt_sin2beta_NP}) and shown by the lower right plot of Fig.~\ref{fig:UTfit}. 
In that scenario one finds  $\Delta M_d^{\rm SM} / \Delta M_s^{\rm SM}$ 
to be much higher than the actual measurement. In order to agree exactly with the experimental central value, one needs a NP contribution to $\Delta M_d / \Delta M_s$ at the level of $-22\%$. Non-universal contributions suppressing $\Delta M_d$ ($C_{B_d} < {1}$) and/or enhancing $\Delta M_s$ ($C_{B_s} >{1}$) could be responsible for this shift as is evident from (\ref{eq:Rt_sin2beta_NP}).
The increased value of $R_t$ that compensates the negative effect of NP in
$\Delta M_d/\Delta M_s$ allows to fit the experimental value of $\epsilon_K$.
This solution is characterized by a large value of $\gamma\simeq 84^\circ$ and $\alpha$
much below $90^\circ$. The latter fact could become problematic for this solution when the determination
of $\alpha$ further improves. 

 {\bf 4.} The value of $\vcb$ is significantly increased to roughly 
           $43.5\cdot 10^{-3}$, which seems rather unlikely.

The first  three NP scenarios characterized by black points in
Fig.~\ref{fig:UTfit} will be clearly distinguished from each other once the
values of $\gamma$ and $R_b$ from tree level decays will be precisely
known. Moreover, if  future measurements of $(R_b,\gamma)$ will select a
point in the $R_b - \gamma$ plane that differs from the black points in
Fig.~\ref{fig:UTfit}, it is likely that NP will simultaneously enter
$\epsilon_K$, $S_{\psi K_S}$ and $\Delta M_d / \Delta M_s$. It will be
interesting to monitor future progress in the $R_b-\gamma$ plane.

Finally, let us mention that the tensions discussed above could 
be in principle somewhat reduced through penguin contributions
to $B\to\psi K_S$ \cite{Ciuchini:2005mg,Faller:2008zc}. However
 a different view has been 
expressed in \cite{Gronau:2008cc}, where such effects have been found
to be negligible.

\subsection{Rare Decays $\mathbf{\kpn}$ and $\mathbf{\klpn}$}
Let us next discuss in more detail two most popular decays among 
rare $K$ decays: $\kpn$ and $\klpn$.
These decays are  theoretically 
very clean and  very sensitive to NP contributions in $Z$ penguin 
diagrams. It is then not surprising that theorists invested over 25 
years to improve the SM prediction and to analyze these decays in 
many extensions of the SM. The most recent predictions that include 
NNLO QCD corrections and
electroweak corrections read
\cite{Buras:2006gb,Brod:2008ss}
\be
Br(\kpn)_{\rm SM}=(8.5\pm0.7)\cdot 10^{-11},
\ee

\be\label{brklpn}
Br(\klpn)_{\rm SM}=(2.8\pm0.6)\cdot 10^{-11},
\ee
where the errors are dominated by parametrical uncertainties, in particular
by the CKM parameters. In the past a sizable uncertainty in $Br(\kpn)$
was due to the charm quark mass. But presently $m_c$ is known to be 
$m_c(m_c)=1.279\pm0.013\gev$ 
\cite{Chetyrkin:2009fv} and this 
uncertainty is basically eliminated. Also very significant progress has
been made in estimating non-perturbative contributions to the charm 
component \cite{Isidori:2005xm} and in the determination of the 
relevant hadronic matrix 
elements from tree level leading $K$ decays \cite{Mescia:2007kn}.
Reviews
of these two decays can be found in 
\cite{Buras:2004uu,Smith:2006qg,Isidori:2007zs,Smith:2009bw,Gorbahn:2009jd}. 

On the experimental side seven events of $\kpn$ decay have been observed 
by E787 and E949 at Brookhaven resulting in 
\cite{Artamonov:2008qb}
\be\label{kpnexp}
Br(\kpn)=(17.3^{+11.5}_{-10.5})\cdot 10^{-11}\,.
\ee
The experimental upper bound on $Br(\klpn)$ is still by more than two orders
of magnitude above the SM value in (\ref{brklpn}) but the present upper bound
from E391a at KEK
\cite{Ahn:2007cd} of $Br(\klpn)\le 6.7\cdot 10^{-8}$ 
should be significantly improved in the coming decade.
Experimental prospects for both decays have been already mentioned in 
connection with  Goal 10 on our list.

Once measured, these decays will provide a very clean determination of the 
angle $\beta$ in the UT as some parametric uncertainties, 
in particular the value
of $\vcb$, cancel out in this determination. This implies one of the 
{\it 
golden relation} of  MFV 
\cite{Buchalla:1994tr,Buras:2001af} that connects $K$ and $B$ physics,
\be\label{gold1}
(\sin2\beta)_{S_{\psi K_S}} = (\sin2\beta)_{\klpn}~, 
\ee 
which can be strongly violated 
in models with new flavour and CP-violating interactions, such as
 the LHT model \cite{Blanke:2006eb,Blanke:2009am} and the RSc  model
analyzed in \cite{Blanke:2008yr}. 

While the test of the relation (\ref{gold1}) in future experiments will
tell us whether some NP disturbs this  MFV correlation, in order to
identify which NP is at work we have to do much more and consider 
other decays and observables.

{
\begin{table}[ht]
\renewcommand{\arraystretch}{1}\setlength{\arraycolsep}{1pt}
\center{\begin{tabular}{|c|c|c|c|c|}
\hline
Model/Observable & $Br(\kpn)$ & $Br(\klpn)$ & $Br(B_s\to\mu^+\mu^-)$ &
$S_{\psi\phi}$\\
\hline
 CMFV  & $20\%$ & $20\%$ & $20\%$ & 0.04 \\
 MFV  & $30\%$ & $30\%$ & $1000\%$ & 0.04 \\
 AC  & $2\%$ & $2\%$ & $1000\%$ & 1.0 \\
 RVV2  & $10\%$ & $10\%$ & $1000\%$ & 0.50 \\
 AKM  & $10\%$ & $10\%$ & $1000\%$ & 0.30 \\
$\delta$LL  & $2\%$ & $2\%$ & $1000\%$ & 0.04 \\
FBMSSM  & $2\%$ & $2\%$ & $1000\%$ & 0.04 \\
GMSSM  & $300\%$ & $500\%$ & $1000\%$ & 1.0 \\
LHT  & $150\%$ & $200\%$ & $30\%$ & 0.30 \\
RSc  & $60\%$ & $150\%$ & $10\%$ & 0.75 \\
\hline 
\end{tabular}  }
\caption {Approximate maximal enhancements for various observables in different 
models of NP. In the case of $S_{\psi\phi}$ we give 
the maximal positive values. The NP models have been  defined in Section 2.4.}
\label{tab:summary}
\renewcommand{\arraystretch}{1.0}
\end{table}
}

To this end let us first list the maximal enhancements of these two branching
ratios in a number of NP scenarios. 
These are given in the second and third column of Table~\ref{tab:summary}, where 
$100\%$ means an enhancement of the branching ratio by a factor of two.
These enhancements in a given NP scenario are consistent with all existing 
data but could be significantly decreased through various correlations when
new observables will be measured.

A striking hierarchy of enhancements is exhibited in this table:
\begin{itemize}
\item
 In the GMSSM still very large enhancements are possible. More modest
 but still large enhancements are possible in the LHT model 
\cite{Blanke:2006eb,Blanke:2009am} and in the RSc model \cite{Blanke:2008yr}.
 In the GMSSM and the LHT model the central  experimental value of 
$Br(\kpn)$ in  (\ref{kpnexp}) can be reproduced. In the RSc model values 
above $15\times 10^{-11}$ are rather unlikely.
\item
 Enhancements of both branching ratios in CMFV and MFV scenarios are small,
 but as the theory is very clean, powerful experiments will be able to
 distinguish 
 these NP scenarios on the basis of these decays one day. Yet, the 
 confirmation of the central value for $Br(\kpn)$ in (\ref{kpnexp})
 with a precision of $10\%$ would certainly tell us that non-MFV interactions
 are at work.
 \item
 The branching ratios for both decays in supersymmetric flavour models 
 considered in subsequent subsections
 are basically indistinguishable from the SM predictions for 
$K\to\pi\nu\bar\nu$ decays, but as we will see soon 
 these models perform quite 
 differently in the $B_s$ system or more explicitly in $b\to s$ 
 transitions, both CP-conserving and CP-violating.
\end{itemize}

\subsection{The VIP's of $\mathbf{B_s}$ Physics: 
$\mathbf{B_{s,d}\to \mu^+\mu^-}$ and $\mathbf{S_{\psi\phi}}$}
 We will move now to discuss  Goals 6 and 4 in more detail. These goals 
are in my 
 opinion the most important goals in quark flavour physics
 until the next EPS11 conference, to be joined later by $K\to\pi\nu\bar\nu$
 decays so that EPS13 will indeed have them all. 
We will first discuss these two goals separately. Subsequently 
 we will have a grand simultaneous 
 look at $S_{\psi\phi}$, $B_s\to\mu^+\mu^-$, 
 $K^+\to\pi^+\nu\bar\nu$ and $\klpn$ that we have already anticipated when 
constructing Table~\ref{tab:summary}.

\subsubsection{$\mathbf{Br(B_{s,d}\to \mu^+\mu^-)}$}
One of the main targets of flavour physics in the coming years will be the
measurement of the branching ratio for the highly suppressed decay 
$B_s\to\mu^+\mu^-$. Hopefully  the even more suppressed decay 
$B_d\to \mu^+\mu^-$ will be discovered as well. These two decays are helicity
suppressed in the SM and CMFV models. Their branching ratios are proportional to the
squares of the corresponding weak decay constants that suffer still from
sizable uncertainties as discussed in the context of  Goal 2. However using
simultaneously the SM expressions for the very well measured mass 
differences $\Delta M_{s,d}$, this uncertainty can be 
eliminated \cite{Buras:2003td}
leaving the uncertainties in the hadronic parameters
 $\hat B_{B_s}$  and $\hat B_{B_d}$
as the only theoretical uncertainty 
in $Br(B_{s,d}\to\mu^+\mu^-)$. As seen in 
(\ref{BBB}) these parameters are already known from
lattice calculations \cite{Lubicz:2008am} with precision of $10\%$ 
and enter the branching ratios linearly.

 Explicitly we have in the SM \cite{Buras:2003td}
\be\label{R2}
Br(B_{q}\to\mu^+\mu^-)
=C\frac{\tau(B_{q})}{\hat B_{B_{q}}}
\frac{Y^2(x_t)}{S(x_t)} 
\Delta M_{q}, \qquad (q=s,d)
\ee
with 
\be
C={6\pi}\frac{\eta_Y^2}{\eta_B}
\left(\frac{\alpha}{4\pi\sin^2\theta_{W}}\right)^2\frac{m_\mu^2}{\mw^2}
=4.39\cdot 10^{-10}.
\ee
$S(x_t)=2.32\pm 0.07$ and $Y(x_t)=0.94\pm0.03$ are the relevant 
top mass dependent
one-loop functions.
More generally we have in CMFV models
\be\label{R2a}
\frac{Br(B_{q}\to\mu\bar\mu)}{\Delta M_{q}}
=4.4 \cdot 10^{-10} \frac{\tau(B_{q})}{\hat B_{q}} F(v),
 \qquad F(v)=\frac{Y^2(v)}{S(v)},
\ee
 with $Y(v)$ and $S(v)$ replacing $Y(x_t)$ and $S(x_t)$ in a given
CMFV model. 
Using these expressions one finds in the SM the rather precise predictions
\be\label{TH}
Br(B_s\to \mu^+\mu^-)= (3.6\pm0.4)\cdot 10^{-9}, \qquad
Br(B_d\to \mu^+\mu^-)= (1.1\pm0.1)\cdot 10^{-10}.
\ee

These predictions should be compared to the $95\%$ C.L. upper limits from 
CDF  \cite{Aaltonen:2007kv} and D0 \cite{Abazov:2007iy} (in parentheses)
\be\label{CDFD0}
Br(B_s\to \mu^+\mu^-)\le 3.3~(5.3)\cdot 10^{-8}, \qquad
Br(B_d\to \mu^+\mu^-)\le 1 \cdot 10^{-8}.
\ee
The numbers given above are updates presented at this conference.
More information is given by Giovanni Punzi.
It is clear from (\ref{TH}) and (\ref{CDFD0}) 
that a lot of room is still left for NP contributions.

Now, irrespectively of large uncertainties in the separate  SM predictions  
for $B_{s,d}\to\mu^+\mu^-$ and $\Delta M_{s,d}$, there exists a rather precise
relation between these observables that can be considered as one of the 
theoretically 
cleanest predictions of CMFV. This 
{\it golden relation} reads \cite{Buras:2003td}
 \be\label{eq:r}
\frac{Br(B_s\to\mu^+\mu^-)}{Br(B_d\to\mu^+\mu^-)}= \frac{\hat
B_{B_d}}{\hat B_{B_s}} \frac{\tau(B_s)}{\tau(B_d)} \frac{\Delta
M_s}{\Delta M_d}\,r\,,
\ee
with $r=1$ in CMFV  models but generally different from unity.
For instance in the LHT model one finds $0.3\le r\le 1.6$   
\cite{Blanke:2006eb,Blanke:2009am},  while in the
RSc model $0.6\le r \le 1.3$ \cite{Blanke:2008yr}. Also in 
supersymmetric models discussed below $r$ can deviate strongly
from unity.

It should be stressed that the ratio 
$\hat B_{B_d}/\hat B_{B_s}=1.00\pm 0.03$ \cite{Lubicz:2008am} 
constitutes the only theoretical  uncertainty in
(\ref{eq:r}). The remaining quantities entering (\ref{eq:r})
 can be obtained directly from experimental
data. The right hand side is already known rather precisely: $32.5\pm 1.7$, 
but it will
still take some time before the left hand side will be known with comparable
precision unless NP enhances both branching ratios by an order of magnitude.
In the latter case one will very likely find 
$r\not=1$ as within CMFV models
such large enhancements of $Br(B_{s,d}\to\mu^+\mu^-)$ are not possible.

\begin{figure}[tbp]
\begin{center}
\includegraphics[width=2.7in]{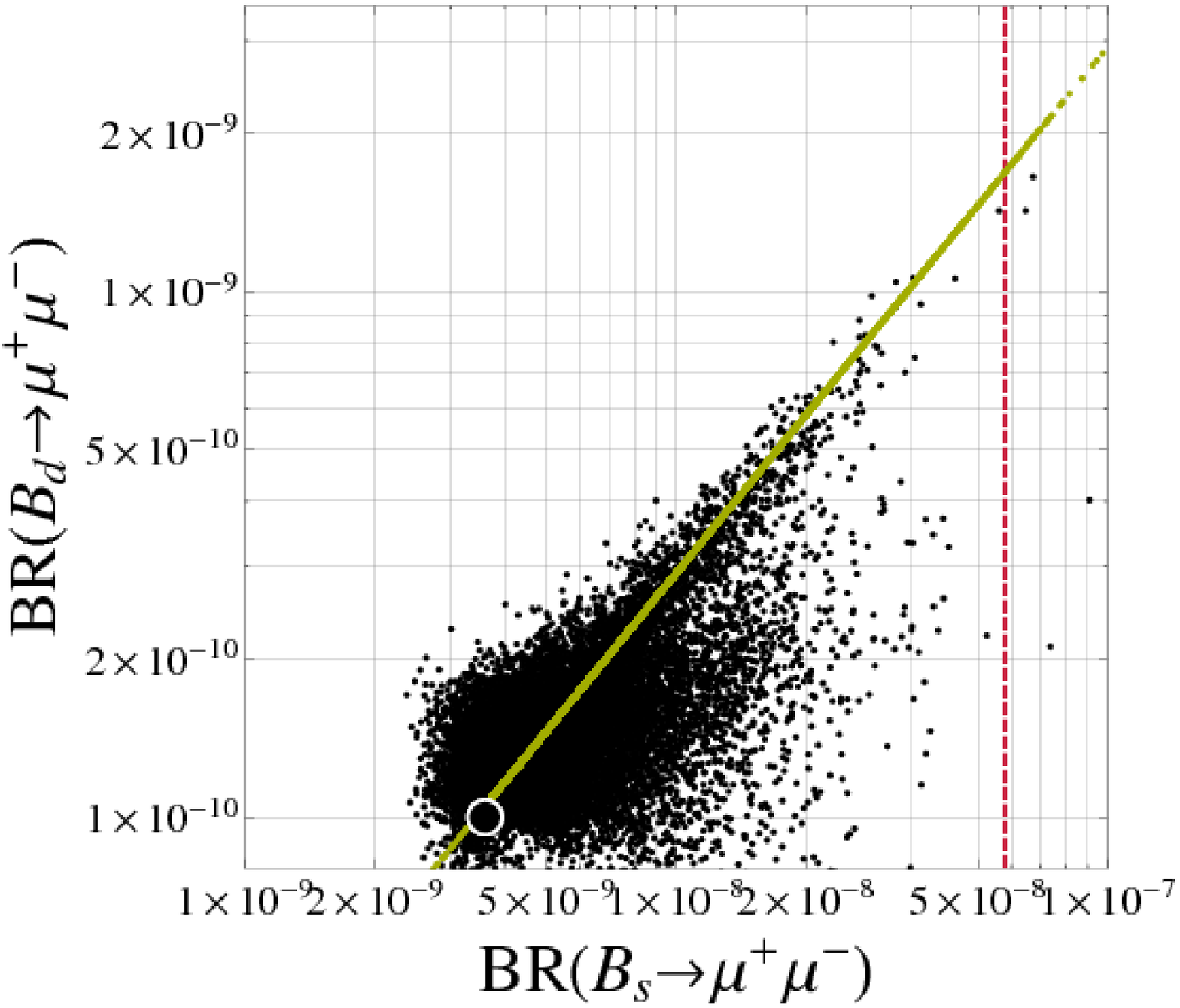}
\includegraphics[width=2.7in]{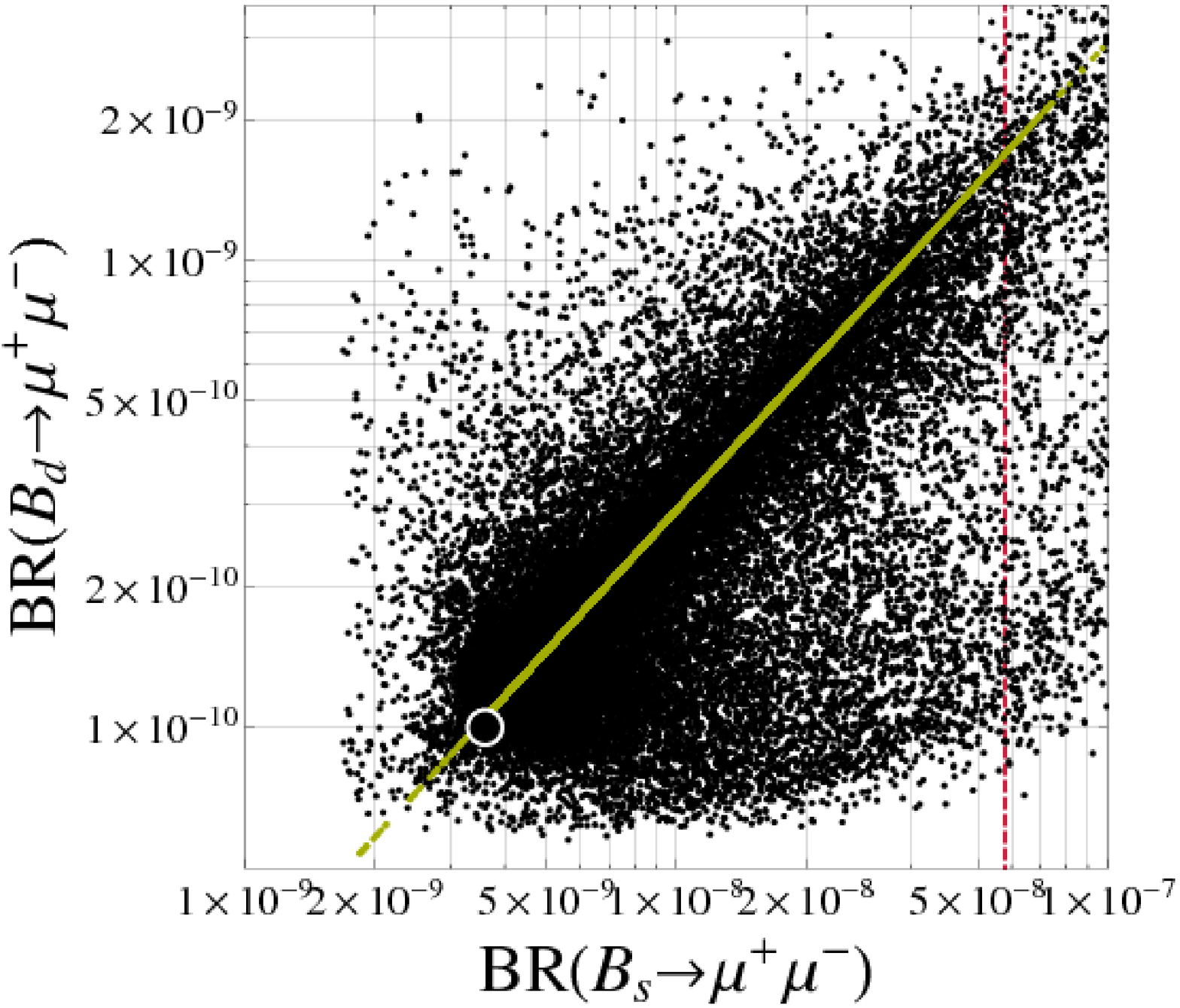}
\end{center}
\caption{\label{Bds} $B_{d,s}\to \mu^+\mu^-$ branching ratios in the
RVV2 model (left) and the $\delta$LL model (right) as obtained 
in \cite{Altmannshofer:2009ne}. }
\end{figure}

Large contributions to the branching ratios in question can come from neutral
scalar exchanges (Higgs penguins) \cite{Choudhury:1998ze,Babu:1999hn}
in which case new scalar operators are generated and the
helicity suppression is lifted. Thus large enhancements of $B_{s,d}\to
\mu^+\mu^-$ are only possible in the models placed in the entries (1,2) and
(2,2) of the flavour matrix in Fig.~\ref{fig:matrix}. The prime example here 
is 
the MSSM at large $\tan\beta$,
in which still in 2002 $Br(B_s\to\mu^+\mu^-)$ could be as large as 
$10^{-6}$. The 
impressive progress by CDF and D0 collaborations, leading to a 
 decrease of the
corresponding upper bound by two orders of magnitude totally excluded this
possibility but there is still  hope that  a clear signal of NP at the 
level of $\ord(10^{-8})$ will be seen
in these decays. We will discuss a number of SUSY predictions
below, where such enhancements are still possible.

In the MSSM with MFV and large $\tan\beta$ there is a strong correlation
between $Br(B_{s,d}\to \mu^+\mu^-)$ and $\Delta M_s$ 
\cite{Buras:2002vd,Buras:2002wq,Isidori:2002qe,Carena:2006ai,Gorbahn:2009pp} implying that an
enhancement of these branching ratios with respect to the SM is correlated
with a suppression of $\Delta M_s$ below the SM value. In fact the MSSM with
MFV was basically the only model that ``predicted'' the suppression of 
$\Delta M_s$ below the SM prediction as seemed to be the case just after
the discovery of the $B^0_s-\bar B^0_s$ mixing. 
Meanwhile the lattice values for
weak decay constants changed and there is no suppression relativ 
to $(\Delta M_s)_{\rm SM}$ seen within theoretical
uncertainties in the data. With the decrease of the experimental 
upper bound on $Br(B_{s,d}\to \mu^+\mu^-)$ also in the MSSM with MFV 
the predicted suppression of $\Delta M_s$ amounts
to at most $10\%$ and it will require a considerable reduction of the lattice
uncertainties in the evaluation of $(\Delta M_s)_{\rm SM}$  before the
correlation in question can be verified or falsified by experiment.
As we will see soon, in the MSSM with non-MFV interactions the correlation 
discussed here is
absent \cite{Altmannshofer:2009ne}. Other analyses of this issue can be
found in \cite{Chankowski:2002wr,Dedes:2002er,Dedes:2008iw} and a review 
on Higgs penguins can be found in \cite{Dedes:2003kp}. Also in models
with hybrid gauge-gravity mediation the MFV-like correlattion in question
can be strongly modified \cite{Hiller:2008sv}.

Looking at $Br(B_{s,d}\to\mu^+\mu^-)$ in CMFV, MFV, LHT, RSc, GMSSM and 
the specific supersymmetric flavour models AC, RVV2, AKM, $\delta$LL and  
FBMSSM
we identify a striking hierarchy of possible enhancements that is,  as seen in 
table~\ref{tab:summary}, 
opposite to the one found in the case of 
$K\to\pi\nu\bar\nu$ decays. An exception to this pattern is GMSSM:
\begin{itemize}
\item
In the GMSSM, SUSY with MFV and all SUSY flavour models 
 $Br(B_{s,d}\to\mu^+\mu^-)$ can still
reach the present experimental bounds because of the presence of 
Higgs penguins that become very important at large $\tan\beta$: a 
$(\tan\beta)^6$ enhancement of the branching ratios is present in this 
case.
\item
In CMFV, the LHT and the RSc only enhancements of $20\%$, $30\%$ and $10\%$ 
are possible \cite{Blanke:2006eb,Blanke:2009am,Blanke:2008yr}
as Higgs penguins are irrelevant here and the Z-penguins in spite 
of non-MFV interactions in the case of the LHT and the RSc do not lift 
the helicity suppression. Moreover the custodial protection of 
left-handed Z couplings in the 
RSc allows only right-handed Z couplings to be relevant and these
cannot do much in this case \cite{Blanke:2008yr}.
\end{itemize}

 Recently a closer look at $Br(B_{s,d}\to\mu^+\mu^-)$ has been made 
 in the context of specific SUSY flavour models such as 
AC, RVV2, AKM, $\delta$LL 
 showing that the measurement of both branching ratios
 $Br(B_{s,d}\to \mu^+\mu^-)$ can not only test the golden MFV relation 
 in (\ref{eq:r}) but also give some insight in different SUSY flavour models.
 We find \cite{Altmannshofer:2009ne}: 
\begin{itemize}
\item
The ratio 
$Br(B_d\to\mu^+\mu^-)/Br(B_s\to\mu^+\mu^-)$ in the AC and RVV2  models is 
dominantly below its CMFV prediction in (\ref{eq:r}) and can be much 
smaller than the latter.
In the AKM model this ratio stays much closer to the MFV value of roughly 
$1/33$ \cite{Buras:2003td,Hurth:2008jc} and can be smaller or larger than 
this value with equal probability.
Still, values of $Br(B_d\to\mu^+\mu^-)$ as high as $1\times 10^{-9}$ are
possible in all these models as $Br(B_s\to\mu^+\mu^-)$ can be strongly 
enhanced. We show this in the case of the RVV2 model in the left plot of
Fig.~\ref{Bds}.
\item
Interestingly,  in the $\delta$LL-models, the ratio 
$Br(B_d\to\mu^+\mu^-)/Br(B_s\to\mu^+\mu^-)$ can not only deviate
significantly from its CMFV value, but in contrast to the models with 
right-handed currents considered by us can also be much 
larger that the MFV value. 
Consequently,
$Br(B_d\to\mu^+\mu^-)$ as high as $(1-2)\times 10^{-9}$ is still 
possible while being consistent with the bounds on all other observables,
in particular the one on $Br(B_s\to\mu^+\mu^-)$. We show this in the 
right plot of Fig.~\ref{Bds}.
\end{itemize}

\begin{figure}[tbp]
\begin{center}
\includegraphics[width=2.7in]{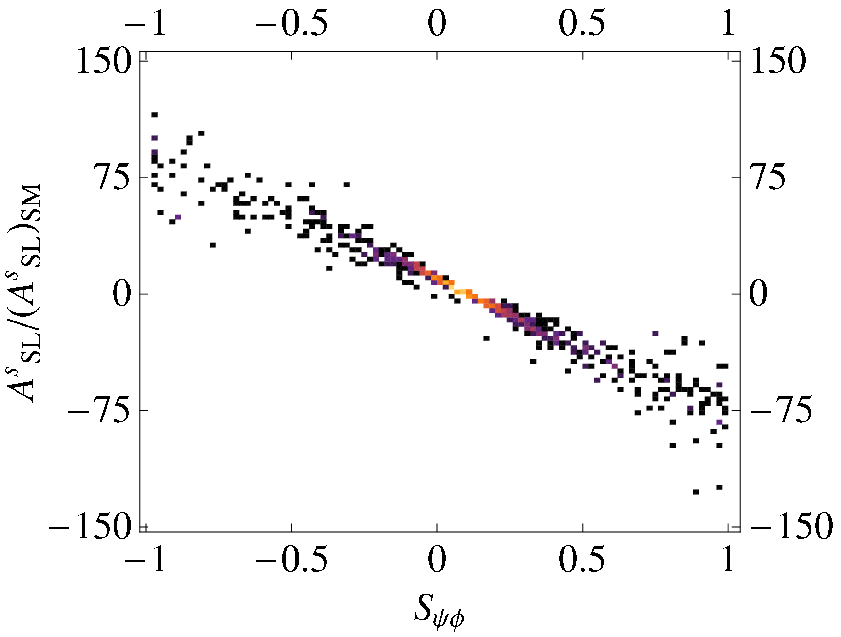}
\includegraphics[width=2.7in]{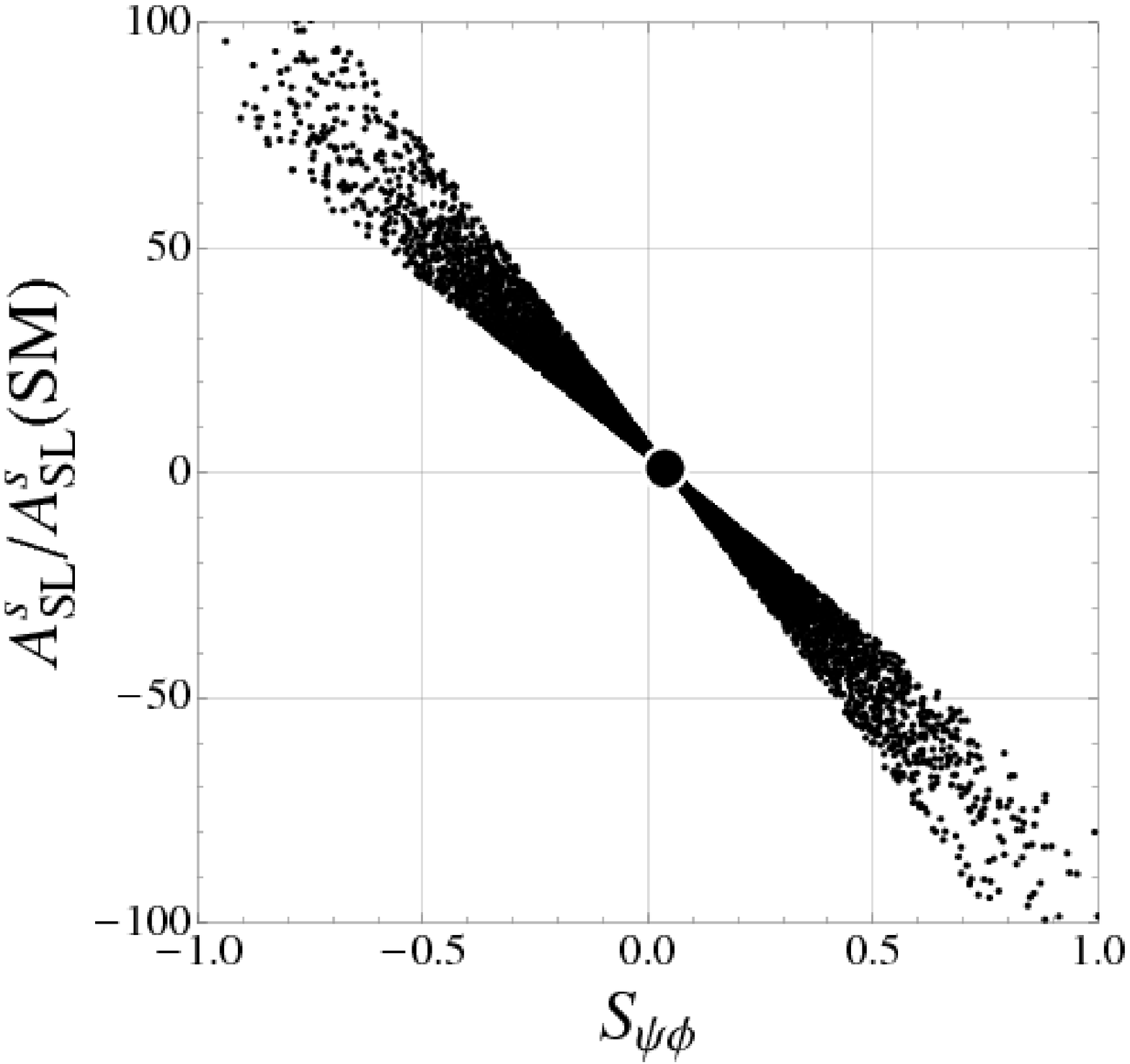}
\end{center}
\caption{\label{ASLS} $A_{SL}^s$ vs. $S_{\psi\phi}$ in the RSc model (left)
\cite{Blanke:2008zb} and in the AC model (right) \cite{Altmannshofer:2009ne}. }
\end{figure}

\subsubsection{The $\mathbf{S_{\psi\phi}}$ Asymmetry}
 The tiny complex phase of the element $V_{ts}$ in the CKM matrix precludes any
sizable CP violating effects in the decays of the $B_s$ mesons within 
the SM and models with MFV.
In 
particular the very clean mixing induced asymmetry $S_{\psi\phi}$ is
predicted to be
\be\label{spp}
(S_{\psi\phi})_{\rm SM}=\sin(2|\beta_s|)\approx 0.04,
\ee
with $-\beta_s$ being the phase of $V_{ts}$. As pointed out in 
\cite{Faller:2008gt}
some hadronic uncertainties not discussed in the past could still be
non-negligible so that values of $S_{\psi\phi}$ as high as $0.1$ 
could not be immidiately considered as signals of NP. However the
same paper proposes various strategies to overcome these difficulties
through additional measurements of different decay channels that will
be available in the coming years.

In the presence of new
physics (\ref{spp}) is modified as follows \cite{Blanke:2006ig},
\be\label{sppnew}
S_{\psi\phi}=\sin(2|\beta_s|-2\phi_{B_s}),
\ee
where $\phi_{B_s}$ is a new phase in $B^0_s-\bar B^0_s$ mixing as 
defined in (\ref{eq:M12d}).

Already in 2006 Lenz and Nierste \cite{Lenz:2006hd}, 
analyzing D0 and CDF data pointed out
some hints for a large phase $\phi_{B_s}$. In 2008 new hints
appeared, emphasized in particular by the UTfit collaboration 
\cite{Bona:2008jn}. 
The most recent messages from CDF and D0 \cite{Brooijmans:2008nt} imply 
 a $2.7\sigma$ deviation
from the SM prediction and we have to wait for higher statistics 
in order to conclude
that NP is at work here \cite{Lenz:2008dp}. 
Explicitly CDF and D0 find \cite{Barberio:2007cr}
\be\label{Spsiphiexp}
S_{\psi\phi}\approx 0.81^{+0.12}_{-0.32}.
\ee
As the central value of the measured $S_{\psi\phi}$ is
around $0.8$, that is one order of magnitude larger than the SM value,
the confirmation of this high value in the future would be a spectacular 
confirmation of non-MFV interactions at work. As demonstrated recently
such large values can easily be found in the RSc model 
\cite{Blanke:2008zb} and the same
comment applies to the GMSSM. The most likely values for 
$S_{\psi\phi}$ in the LHT do not exceed 0.3 \cite{Blanke:2009am} 
and finding this asymmetry
as high as 0.4 would be in favour of the RSc and the GMSSM. Similarly 
the supersymmetric flavour models with significant right-handed currents
(AC, RVV2, AKM) provide sizable enhancements. Here the double Higgs penguin 
 contributing to $M_{12}^s$
is at work. The following hierarchy 
in maximal values is found (see Table~\ref{tab:summary})
\begin{equation}\label{eq:hierarchy}
(S_{\psi\phi})_{\rm LHT}^{\rm max}\approx 
(S_{\psi\phi})_{\rm AKM}^{\rm max}
< (S_{\psi\phi})_{\rm RVV2}^{\rm max}
< (S_{\psi\phi})_{\rm RSc}^{\rm max}
\approx (S_{\psi\phi})_{\rm AC}^{\rm max}.
\end{equation}
On the other hand $S_{\psi\phi}$ in the flavour models with only left-handed 
currents 
and in the FBMSSM is SM-like.

\begin{figure}[tbp]
\begin{center}
\includegraphics[width=2.7in]{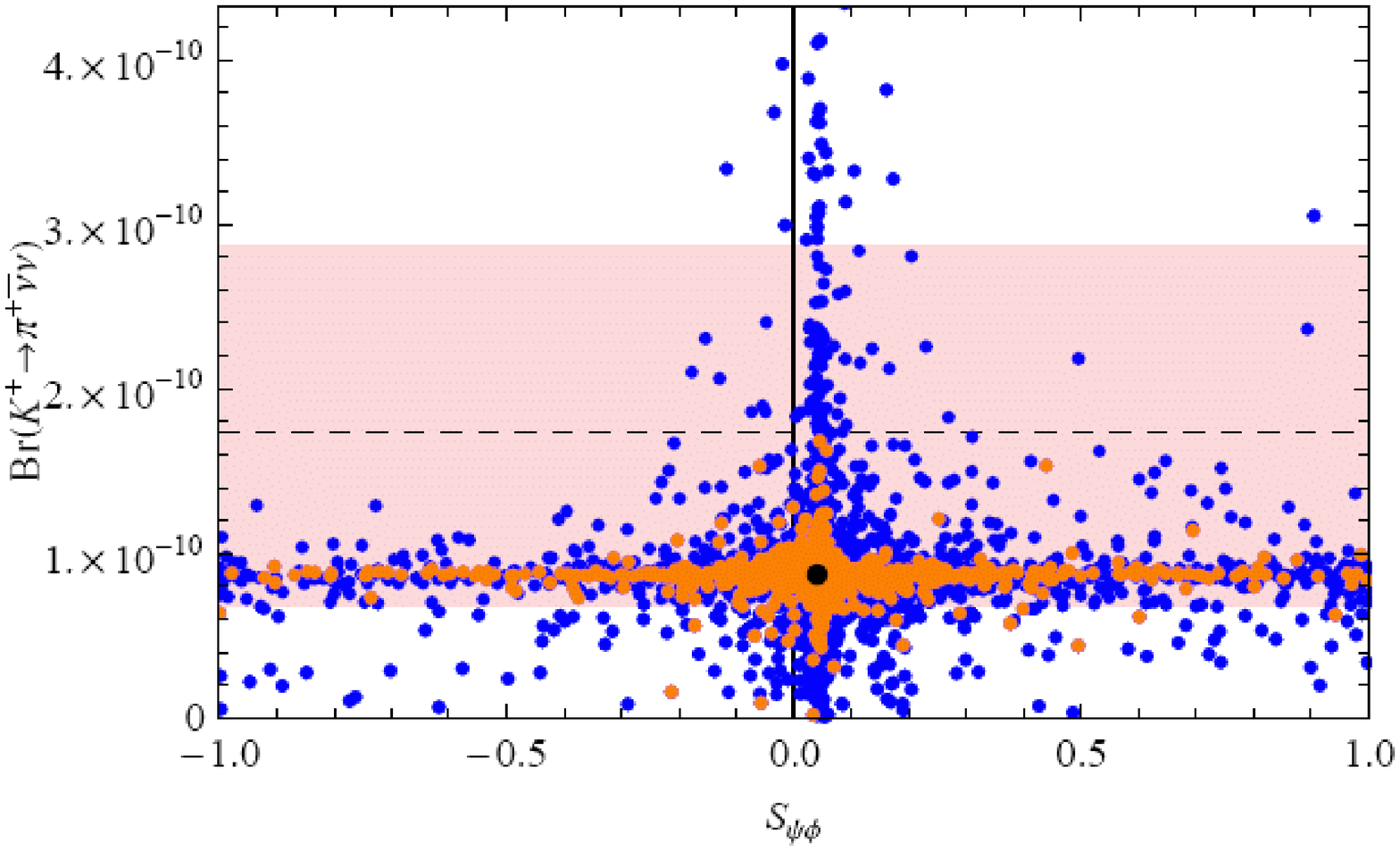}
\includegraphics[width=2.7in]{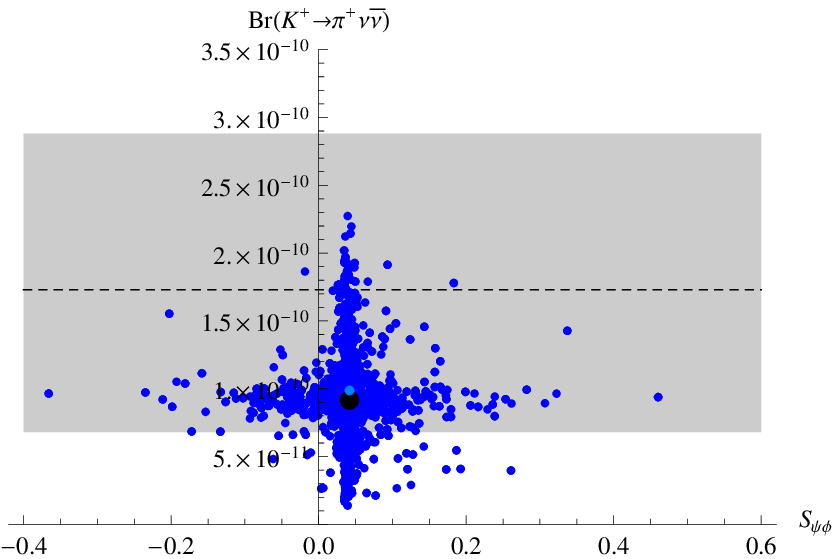}
\end{center}
\caption{\label{KPSP} $Br(\kpn)$ vs. $S_{\psi\phi}$ in the RSc model 
(left) \cite{Blanke:2008yr} and in the LHT model (right) \cite{Blanke:2009am}.
 }
\end{figure}

Clearly a sizable $S_{\psi\phi}$ is not the only manifestation of CP
violation in the $B_s$ system but presently it is the most prominent one
as it can be measured accurately at LHCb, it is theoretically rather clean
and the leftover uncertainties could be further decreased using the
strategies in \cite{Faller:2008gt}. 
In Fig.~\ref{ASLS} we show the correlation between the semi-leptonic 
asymmetry $A_{SL}^s$ and 
 $S_{\psi\phi}$    in the RSc and AC models. This correlation is basically
model independent \cite{Ligeti:2006pm} 
and shows that in any model in which $S_{\psi\phi}$ 
deviates significantly from its SM value, also  $A_{SL}^s$ will be 
very much enhanced.
Other implications of a large $S_{\psi\phi}$ in the context of concrete models 
will be discussed below.

\subsection{Correlations between 
$\mathbf{\kpn}$, $\mathbf{\klpn}$, $\mathbf{B_s\to\mu^+\mu^-}$ 
and $\mathbf{S_{\psi\phi}}$}
In  Table~\ref{tab:summary} we collect the 
largest possible enhancements for the 
corresponding branching ratios and $S_{\psi\phi}$ in
various extensions of the SM discussed in this talk. It is evident that 
if we knew already the values of these four observables that are 
given to us by nature,
we could already make a clear distinction between certain scenarios 
provided the deviations from the SM would be large.

\begin{figure}[thbp]
\begin{center}
\includegraphics[width=2.7in]{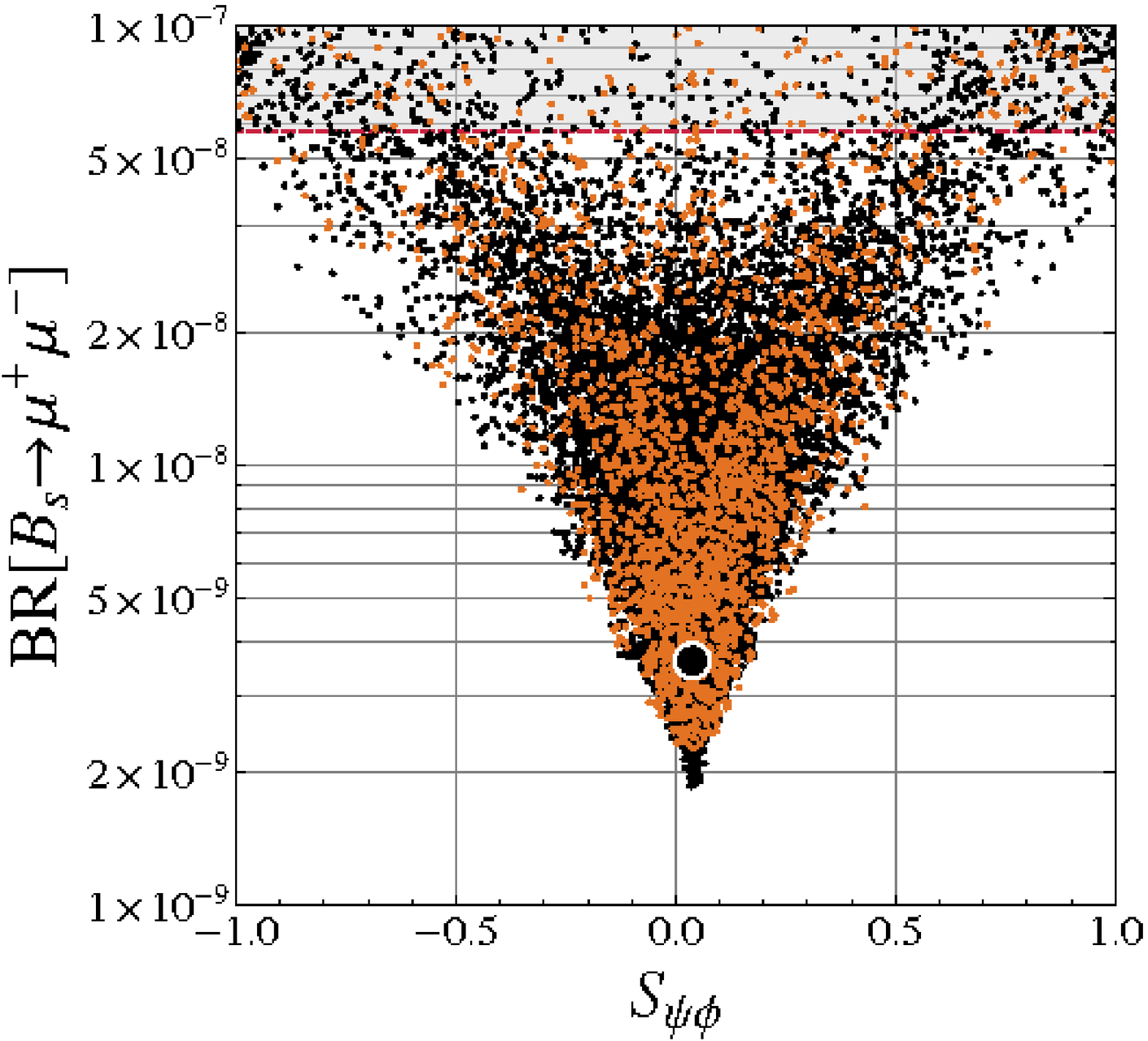}
\includegraphics[width=2.7in]{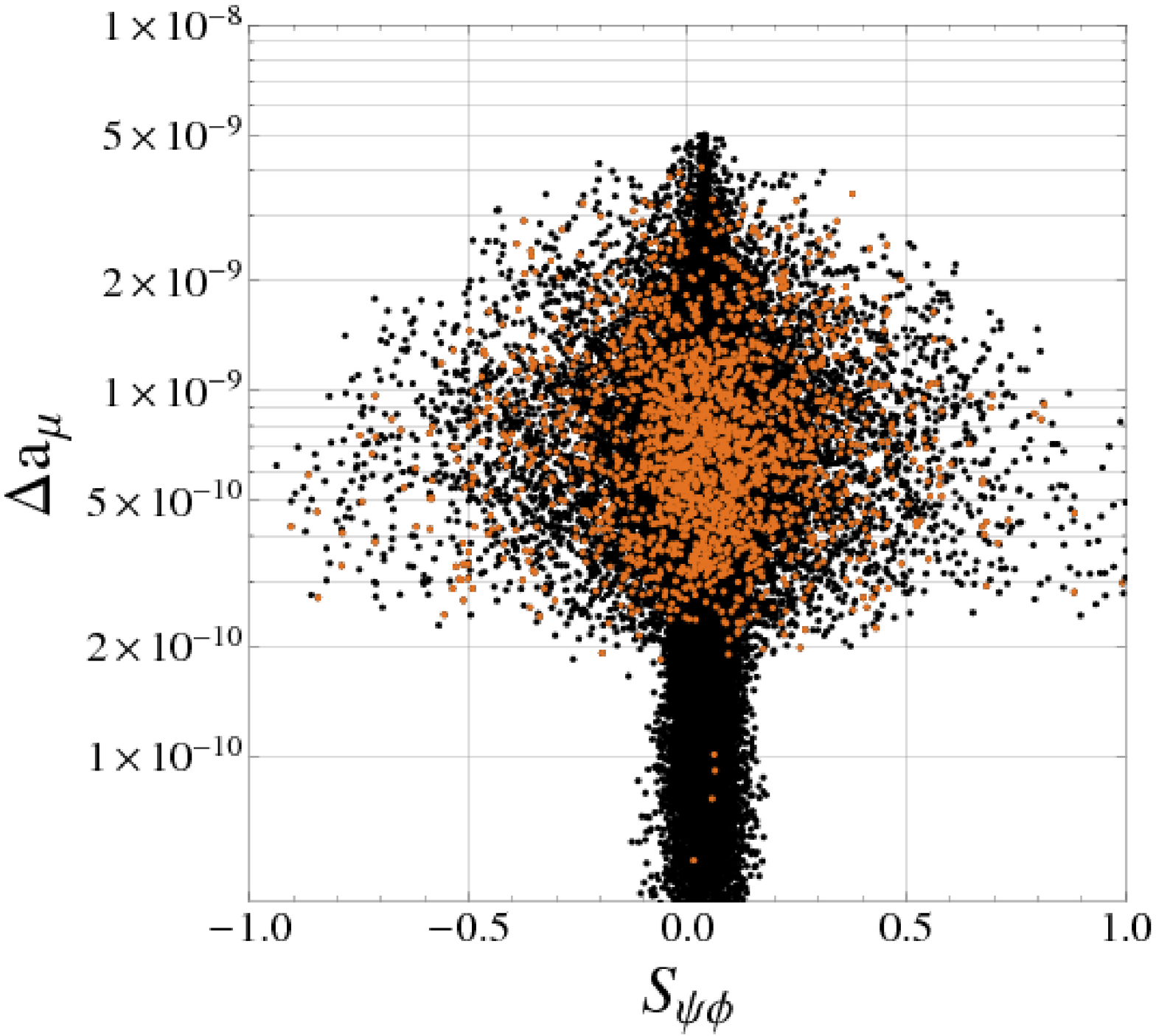}
\end{center}
\caption{\label{AC} $Br(B_{s}\to \mu^+\mu^-)$ vs. $S_{\psi\phi}$ (left)
and $\Delta a_\mu$ vs. $S_{\psi\phi}$ (right) in the AC model as obtained 
in \cite{Altmannshofer:2009ne}. }
\end{figure}

\begin{figure}[thbp]
\begin{center}
\includegraphics[width=2.7in]{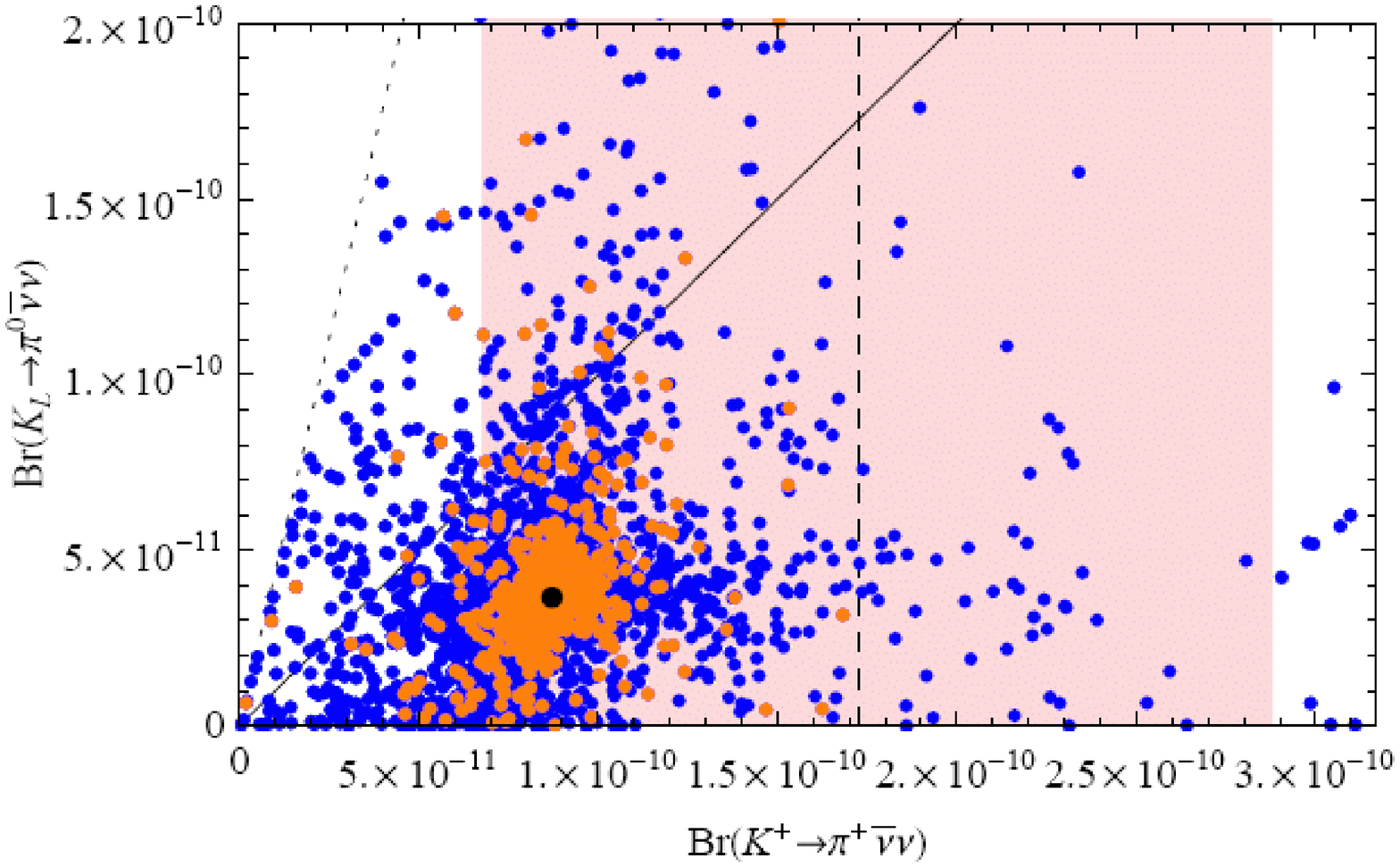}
\includegraphics[width=2.7in]{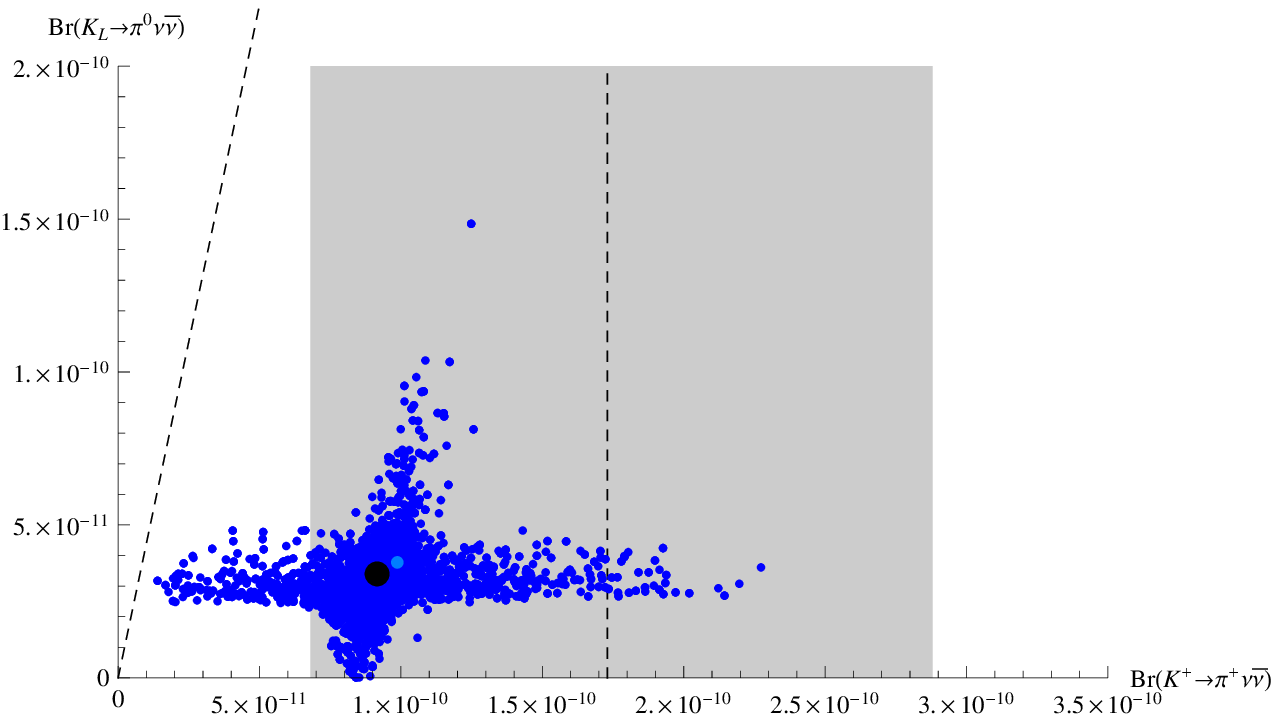}
\end{center}
\caption{\label{KNKP} $Br(\klpn)$ vs. $Br(\kpn)$ in the RSc model 
(left) \cite{Blanke:2008yr} and in the LHT model (right) \cite{Blanke:2009am}.
 }
\end{figure}

This table does not take into account possible correlations between 
these four observables and it is important to list some of them:
\begin{itemize}
\item 
Simultaneous enhancements of $S_{\psi\phi}$ and of
$Br(K\to\pi\nu\bar\nu)$ in the LHT and the RSc scenario are 
rather unlikely \cite{Blanke:2009am,Blanke:2008yr}. This feature
is more pronounced in the RSc model. We show this correlation 
 in Fig.~\ref{KPSP}.
\item
On the contrary the desire to explain the $S_{\psi\phi}$ anomaly within 
the supersymmetric flavour models with
right-handed  currents implies, in the case of the AC and  AKM models,
values of
$Br(B_s\to\mu^+\mu^-)$ as high as several $10^{-8}$. This are 
very exciting news 
for the CDF, D0 and LHCb experiments! In the 
RVV2 model such values are also possible but not necessarily implied
by the large value of $S_{\psi\phi}$. We show one example of this 
spectacular correlation for the case of the AC model in the left plot 
of Fig.~\ref{AC}. 
\item
While in the case of the LHT model some definite correlations 
between $Br(\klpn)$ and $Br(\kpn)$ can be seen \cite{Blanke:2009am}, no such correlations are
found in the case of the RSc model \cite{Blanke:2008yr}, although in both models the enhancements
of the two branching ratios can take place simultaneously. We show this 
 feature in Fig.~\ref{KNKP}.
Some insights in this different behaviour have been recently provided in
\cite{Blanke:2009pq}.
\end{itemize}
More correlations in all these models can be found in the papers 
quoted above but I think the first two on the list above are the most 
interesting in the quark flavour sector. Certainly a precise measurement 
of $S_{\psi\phi}$, in particular if $S_{\psi\phi}$ will be found to be much 
larger than its SM value, will have an important impact on the 
models discussed 
here.

\subsection{The Correlation between the $\mathbf{S_{\psi\phi}}$ and  
$\mathbf{S_{\phi K_S}}$ Anomalies}
Before leaving  quark flavour physics let me return for a moment
to the $S_{\phi K_S}$ anomaly in (\ref{spK}) and discuss it together 
with the $S_{\psi\phi}$ anomaly. These  anomalies can be explained
simultaneously
in the GMSSM but the situation is more interesting  in supersymmetric 
flavour (SF) models.

Indeed the SUSY flavour  models with right-handed currents (AC, RVV2, AKM) 
and those with exclusively left-handed currents ($\delta LL$)
can be globally distinguished by the values of the CP-asymmetries 
$S_{\psi\phi}$ and $S_{\phi K_S}$ with the following important result: 
none of
the models considered by us in \cite{Altmannshofer:2009ne}
can simultaneously explain the $S_{\psi\phi}$ and
$S_{\phi K_S}$ anomalies observed in the data. 
In the models with right-handed currents,
$S_{\psi\phi}$ can naturally be much larger than its SM value, while 
$S_{\phi K_S}$ remains either SM-like or its correlation with $S_{\psi\phi}$ 
is inconsistent with the data. 
On the contrary, in the models with left-handed currents only,
$S_{\psi\phi}$ remains SM-like, while the  $S_{\phi K_S}$  anomaly can easily
be solved. Thus already  precise measurements of 
$S_{\psi\phi}$ and $S_{\phi K_S}$ in the near future 
will select one of these two classes of 
models, if any. 

We will still have something to say about the correlation of these 
two anomalies with observables in the lepton sector in the context 
of the SF models in question.

\subsection{Lepton Flavour Violation, EDM's and $\mathbf{(g-2)_\mu}$}

 Let us finally discuss some additional aspects of Goals 16-18 on our list 
 for the next  decade. In \cite{Altmannshofer:2009ne} we have also performed  a very detailed 
 analysis of LFV, EDM's and 
 of $(g-2)_\mu$ in the supersymmetric flavour models AC, RVV2, AKM and 
$\delta$LL.
 Particular emphasis has been put on 
 correlations between these observables in each of these models and their 
 correlation with flavour observables in the quark sector discussed 
 exclusively in this section until now. Let us just list the 
 most striking results of this study keeping in mind that the models with
 right-handed currents (AC, RVV2, AKM) have the potential to explain the
 $S_{\psi\phi}$ anomaly while the  $\delta$LL model could explain 
the $S_{\phi K_S}$ 
 anomaly. Here we go.

{\bf 1.}
The desire to explain the $S_{\psi\phi}$ anomaly within the models with
right-handed currents automatically  implies a
solution to the $(g-2)_\mu$ anomaly. We illustrate this for the AC model 
in the right plot 
of Fig.~\ref{AC}.

\begin{figure}[thbp]
\begin{center}
\includegraphics[width=2.7in]{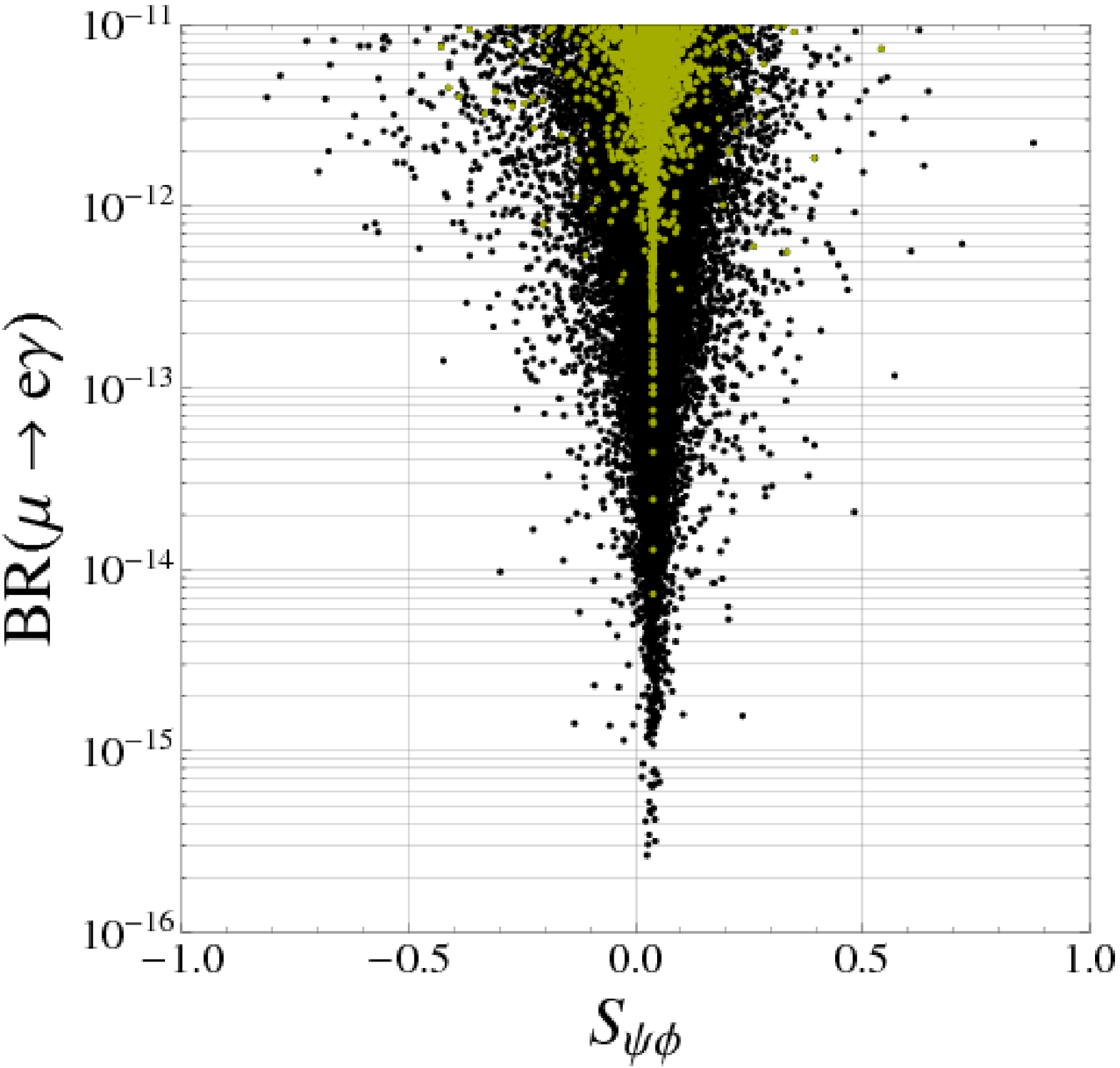}
\includegraphics[width=2.7in]{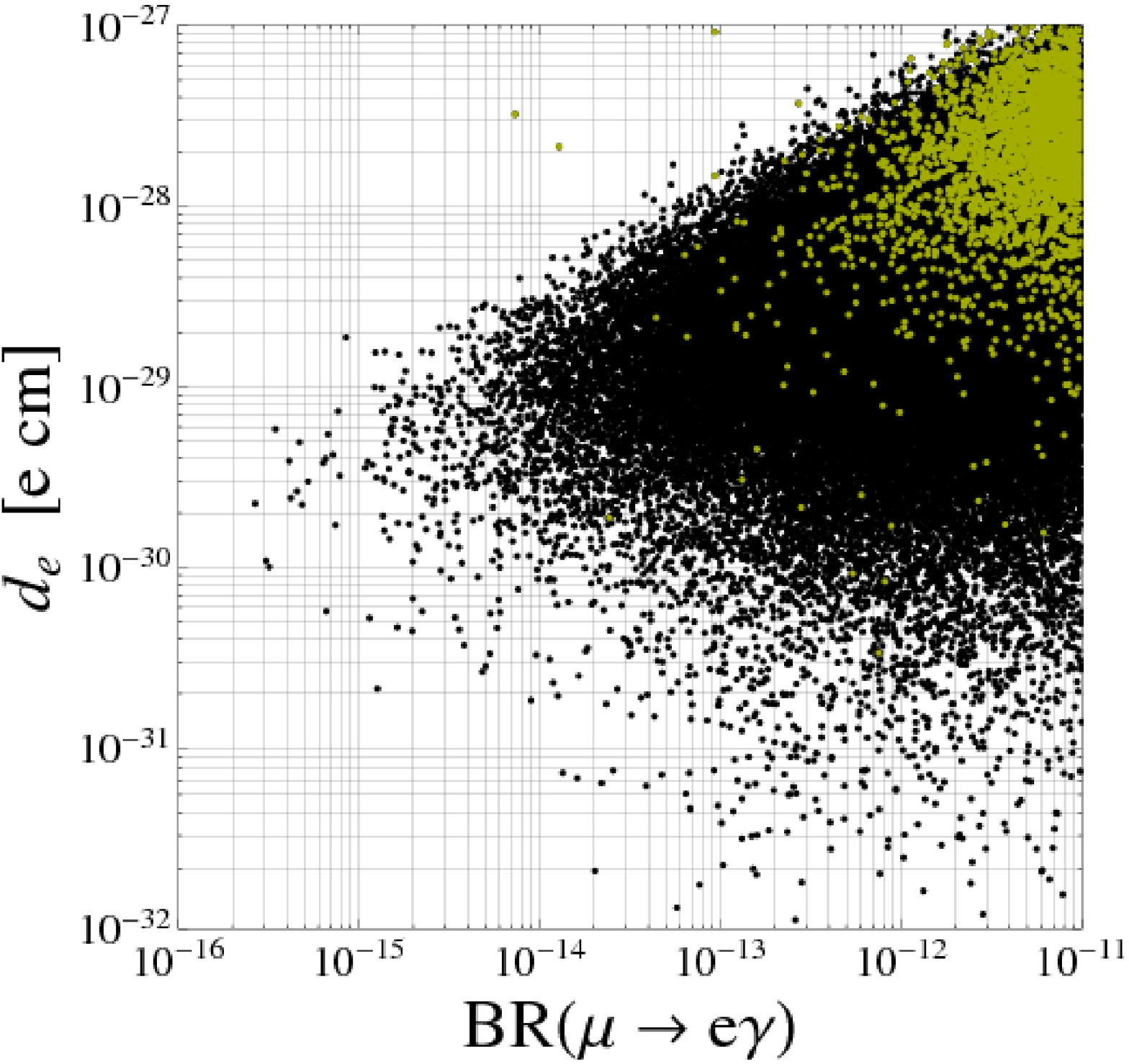}
\end{center}
\caption{\label{RVV2} $Br(\mu\to e\gamma)$ vs. $S_{\psi\phi}$ (left)
and $d_e$ vs. $Br(\mu\to e\gamma)$ (right) in the RVV2 model as obtained 
in \cite{Altmannshofer:2009ne}. The green points explain the $(g-2)_\mu$
anomaly at $95\%$ C.L., i.e. $\Delta a_\mu\ge 1\times 10^{-9}$. }
\end{figure}

{\bf 2.}
In the RVV2 and the AKM models, a large value of $S_{\psi\phi}$  combined with
the desire to explain the $(g-2)_\mu$ anomaly implies 
$Br(\mu\to e\gamma)$ in 
the reach of the MEG experiment.  In the case of the RVV2 model,
$d_e\ge 10^{-29}$ e cm. is predicted, while in the AKM model it is typically
smaller.
 Moreover, in the case of the RVV2 model, 
$Br(\tau\to\mu\gamma)\ge 10^{-9}$ is then
 in the reach of Super-B machines, while this is not the case in the AKM model.
 Some of these results are illustrated in Fig.~\ref{RVV2}.

{\bf 3.} The hadronic EDM's represent very sensitive probes of SUSY flavour 
models with right-handed
currents. In the AC model, large values for the neutron EDM might be easily 
generated by both the
up- and strange-quark (C)EDM. In the former case, visible CP-violating
 effects in 
$D^0-\bar D^0$ mixing
are also expected while in the latter case large CP-violating  effects in the
$B_s$ system are unavoidable.
The RVV2 and AKM models predict values for the down-quark (C)EDM and, 
hence for the neutron
EDM, above the $\approx 10^{-28}e$~cm. level 
if a large $S_{\psi\phi}$
is generated. All the above models predict a 
large strange-quark (C)EDM, hence, a reliable
knowledge of its contribution to the hadronic EDM's 
by means of lattice QCD techniques would
be of the utmost importance to probe or to falsify flavour models 
embedded in a SUSY framework.

\begin{figure}[htbp]
\begin{center}
\includegraphics[width=2.7in]{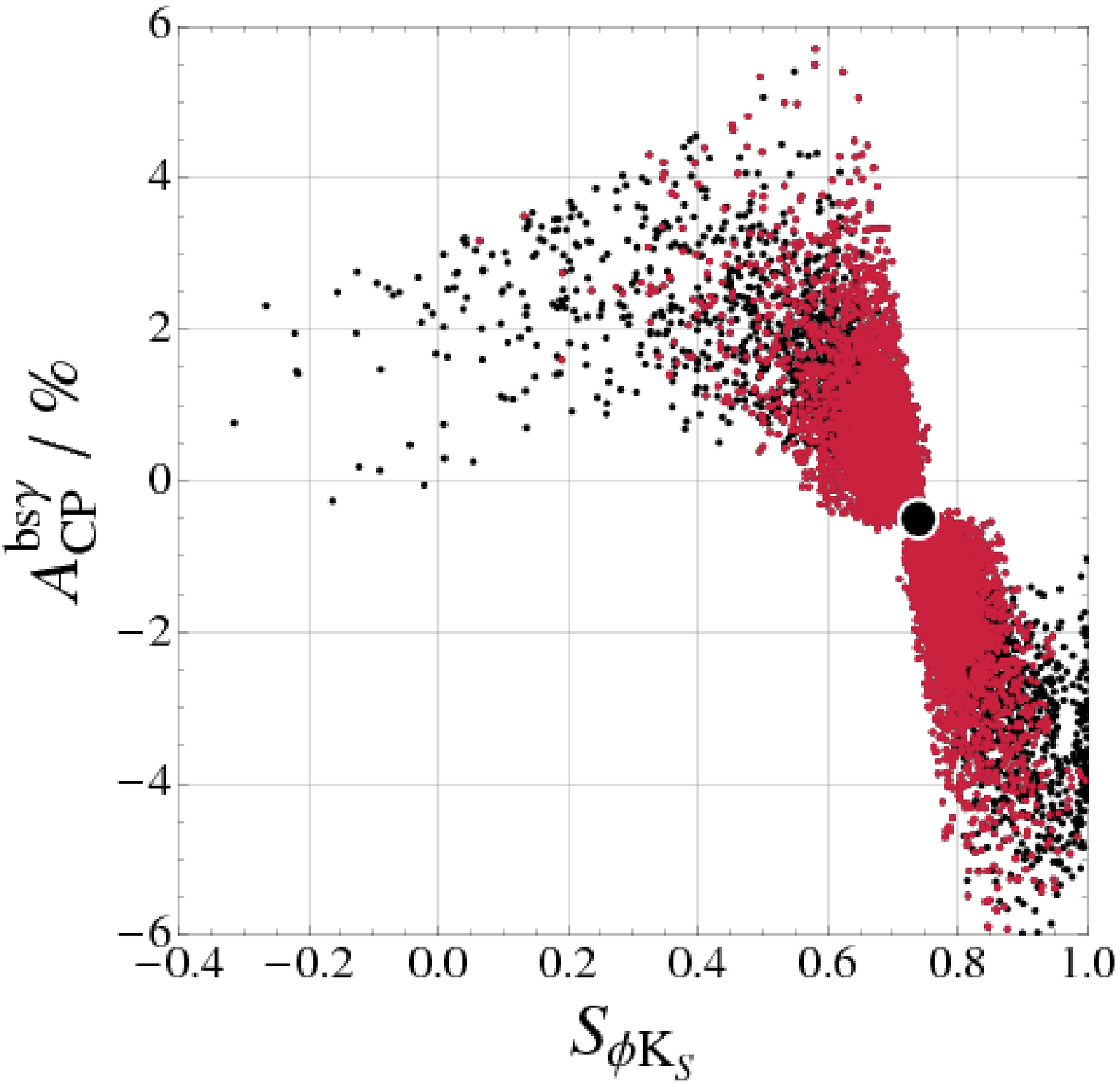}
\includegraphics[width=2.7in]{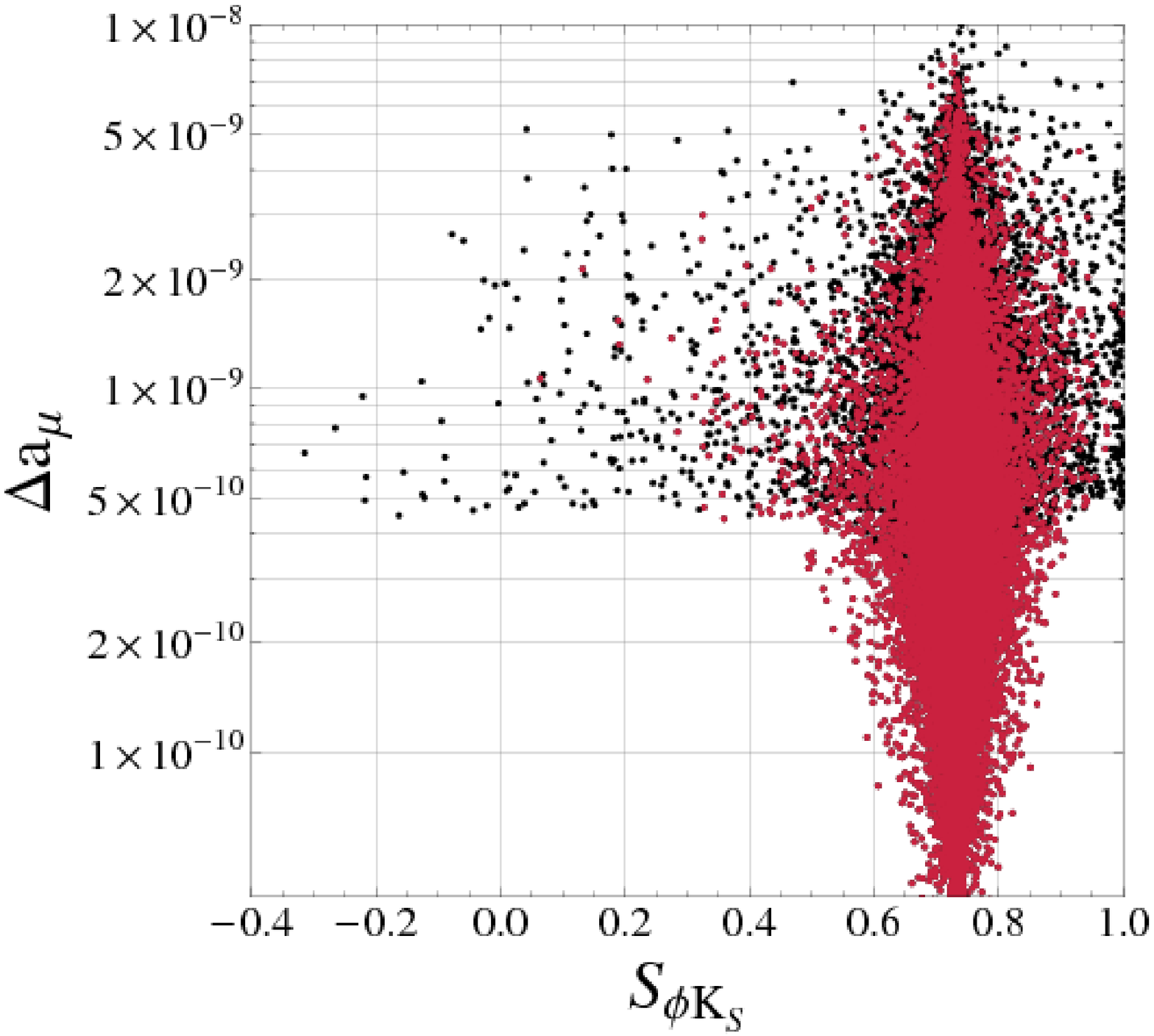}
\end{center}
\caption{\label{dLL} $A_{\rm CP}^{bs\gamma}$ vs. $S_{\phi K_S}$ (left)
and $\Delta a_\mu$ vs. $S_{\phi K_S}$ (right) in the $\delta$LL model as 
obtained 
in \cite{Altmannshofer:2009ne}. The red points satisfy 
$Br(B_s\to\mu^+\mu^-)\le 6 \times 10^{-9}$.}
\end{figure}

{\bf 4.}
In the supersymmetric models with exclusively left-handed currents ($\delta$LL), 
the desire to explain
the $S_{\phi K_S}$ anomaly implies automatically a solution to the 
$(g-2)_\mu$ anomaly and the direct CP asymmetry in $b\to s\gamma$ much
larger than its SM value. We illustrate this in Fig.~\ref{dLL}.
Similar results are found in the 
FBMSSM \cite{Altmannshofer:2008hc}.
This is in contrast to the models with right-handed  currents
where the $A_{\rm CP}^{bs\gamma}$ asymmetry remains SM-like.

\subsection{Testing GUT Models with Rare B Decays}
Next we would like to stress the power of the complex of rare $B$ decays 
 $B\to X_s\gamma$, $B\to X_sl^+l^-$, 
$B_{s,d}\to\mu^+\mu^-$ and $B^+\to \tau^+\nu_\tau$ in testing NP models.
Many analyses of
this type can be found in the literature. Here I would like to mention
only
the analysis of a very interesting $SO(10)$-GUT model of Dermisek and Raby
\cite{Dermisek:2005ij} which gives a successful description of quark and lepton masses, of the
PMNS matrix and of all elements of the CKM matrix except possibly for 
$|V_{ub}|$ that is found to be $3.2\cdot 10^{-3}$, definitely a bit
too low. Yet as shown in \cite{Albrecht:2007ii},
this model fails to describe simultaneously 
the data on the rare decays in question with supersymmetric 
 particles in the reach 
of the LHC. This is mainly due to  $\tan\beta=50$
required in this model. It can be shown that this is a problem of most 
GUTs with Yukawa unification \cite{Altmannshofer:2008vr}. 
Possible solutions to this problem have been 
suggested in that paper. This discussion demonstrates that flavour physics can have
a significant impact not only on physics at the LHC scales but also 
indirectly for
much shorter scales connected with GUT's.

\subsection{A DNA-Flavour Test of New Physics Models}\label{sec:dna}

We have seen above that the patterns of flavour
violation found in various extensions of the SM differed from model to model, 
thereby allowing
in the future to find out which of the models discussed by us, if any, can survive the future
measurements. Undoubtedly, the correlations between various observables that are often
characteristic for a given model will be of the utmost importance in these tests.

In Table~\ref{tab:DNA}, taken from \cite{Altmannshofer:2009ne},
 a summary of the potential size of 
deviations from the SM results
allowed for a large number of observables, considered in that paper and 
here, has been presented, taking into account  all existing constraints
from other observables.
This table can be considered as the collection of the DNA's for various models.
These DNA's will be modified as new experimental data will be availabe and in certain
cases we will be able to declare certain models to be disfavoured or even
ruled out. It should be emphasized that
in constructing the table we did not take into account possible correlations 
among
the observables listed there. We have seen that in some models it is not possible to
obtain large effects simultaneously for certain pairs or sets of observables and
consequently future measurements of a few observables considered in that 
table
will have an impact on the patterns shown there. It will be interesting to
monitor the changes in this table when  future experiments will provide 
new results.

\section{Final Messages and Five Big Questions}

 In our search for a more fundamental theory we need to improve our 
 understanding of  flavour physics.
 The study of flavour physics in conjuction with direct collider searches
 for new physics, with electroweak precision tests and cosmological
 investigations will without any doubt lead one day to a NSM. Whether this will
 happen in 2026 or only in 2046 it is not clear at present. Afterall, 
 35 years have passed since the completion of the present SM and no fully
 convincing  candidate for the NSM exists in the literature. On the
 other hand in view of presently running and upcoming experiments, 
the next decade
 could be like 1970's in which practically every year a new important
 discovery has been made. Even if by 2026 a NSM may not exist yet, it
 is conceivable that we will be able to answer the following crucial
 questions by then:
\begin{itemize}
\item
 Are there any fundamental scalars with masses $M_s\le 1\tev$?
\item
 Are there any new fundamental fermions like vector-like fermions or
 the 4th generation of quarks and leptons?
\item
 Are there any new gauge bosons leading to new forces at very short
 distance scales and an extended gauge group?
\item
 What are the precise patterns of interactions between the gauge bosons,
 fermions and scalars with respect to flavour and CP Violation?
\item
 Can the answers to these four questions help us in understanding the BAU
 and other fundamental cosmological  questions?
\end{itemize}

\definecolor{green1}{rgb}{0.06,0.66,0.06}
\definecolor{orange1}{rgb}{0.98,0.60,0.07}
\newcommand{\green}{{\color{green1}$\bigstar$}}
\newcommand{\orange}{{\color{orange1}\LARGE \protect\raisebox{-0.1em}{$\bullet$}}}
\newcommand{\red}{{\color{red}\small \protect\raisebox{-0.05em}{$\blacksquare$}}}
\newcommand{\three}{{\color{red}$\bigstar\bigstar\bigstar$}}
\newcommand{\two}{{\color{blue}$\bigstar\bigstar$}}
\newcommand{\one}{{\color{black}$\bigstar$}}
%
\begin{table}[t]
\addtolength{\arraycolsep}{4pt}
\renewcommand{\arraystretch}{1.5}
\centering
\begin{tabular}{|l|c|c|c|c|c|c|c|}
\hline
&  AC & RVV2 & AKM  & $\delta$LL & FBMSSM & LHT & RS
\\
\hline\hline
$D^0-\bar D^0$& \three & \one & \one & \one & \one & \three & ?
\\
\hline
$\epsilon_K$& \one & \three & \three & \one & \one & \two & \three 
\\
\hline
$ S_{\psi\phi}$ & \three & \three & \three & \one & \one & \three & \three 
\\
\hline\hline
$S_{\phi K_S}$ & \three & \two & \one & \three & \three & \one & ? \\
\hline
$A_{\rm CP}\left(B\rightarrow X_s\gamma\right)$ & \one & \one & \one & \three & \three & \one & ?
\\
\hline
$A_{7,8}(B\to K^*\mu^+\mu^-)$ & \one & \one & \one & \three & \three & \two & ?
\\
\hline
$A_{9}(B\to K^*\mu^+\mu^-)$ & \one & \one & \one & \one & \one & \one & ? 
\\
\hline
$B\to K^{(*)}\nu\bar\nu$  & \one & \one & \one & \one & \one & \one & \one 
\\
\hline
$B_s\rightarrow\mu^+\mu^-$ & \three & \three & \three & \three & \three & \one & \one
\\
\hline
$K^+\rightarrow\pi^+\nu\bar\nu$ & \one & \one & \one & \one & \one & \three & \three 
\\
\hline
$K_L\rightarrow\pi^0\nu\bar\nu$ & \one & \one & \one & \one & \one & \three & \three
\\
\hline
$\mu\rightarrow e\gamma$& \three & \three & \three & \three & \three & \three & \three \\
\hline\hline
$d_n$& \three & \three & \three & \two & \three & \one & \three
\\
\hline
$d_e$& \three & \three & \two & \one & \three & \one & \three
\\
\hline
$\left(g-2\right)_\mu$& \three & \three & \two & \three & \three & \one & \two
\\
\hline

\end{tabular}
\renewcommand{\arraystretch}{1}
\caption{\small
``DNA'' of flavour physics effects \cite{Altmannshofer:2009ne} for the most interesting observables in a selection of SUSY
and non-SUSY models. \three\ signals large effects, \two\ visible but small effects and \one\
implies that the given model does not predict sizable effects in that observable.}
\label{tab:DNA}
\end{table}

 There are of course many other profound questions \cite{Quigg:2009vq} 
related to grand 
 unification, gravity and string theory and to other aspects of 
 elementary particle physics and cosmology but from my point of view I
 would really be happy if in 2026 satisfactory answers to the five
 questions posed above were available.
 
 In this review written at the advent of the LHC era to which also 
 several low energy precision machines belong, I wanted to emphasize that
 many observables in the quark and lepton flavour sectors have not 
 been measured yet or are only poorly known and that flavour physics 
 only now enters the precision era. Indeed, spectacular 
 deviations from the SM and MFV expectations are still possible in 
 flavour physics.
 The interplay of the expected deviations with direct searches at
 Tevatron, LHC and later at ILC will be most interesting.

 In particular I emphasized the role of correlations between various 
 observables in our search for the fundamental theory of flavour. 
 These correlations and hopefully new discoveries, both in flavour 
 physics and in direct searches for NP will pave the road to the New 
 Standard Model.

~

{\bf Acknowledgments}
I would like to thank the organizers of EPS09 for inviting me to give this 
talk at  such a well organized and interesting conference. 
I would  like to thank all my collaborators for a wonderful
time we spent together exploring different avenues beyond the Standard
Model. Special thanks go to Bj\"orn Duling for invaluable comments on 
the manuscript and to Wolfgang Altmannshofer and Monika Blanke 
for helping me at various stages of this writeup.
This research was partially supported by the Deutsche
Forschungsgemeinschaft (DFG) under contract BU 706/2-1, the DFG Cluster of
Excellence `Origin and Structure of the Universe' and by the German
Bundesministerium f{\"u}r Bildung und Forschung under contract 05HT6WOA.

\end{document}